%% file: chaty_igrj16465-final.tex
\documentclass[printer]{aa} 
\usepackage{natbib,twoopt}
\usepackage{amsmath}
\usepackage{setspace}
\usepackage{graphicx}
\usepackage{multirow}
\usepackage{lmodern}
\usepackage{wrapfig}
\usepackage{pdfpages}
\usepackage{fixltx2e}
\usepackage[capposition=top]{float row}
\input{symbols}
\bibpunct{(}{)}{;}{a}{}{,} 
\makeatletter
\newcommandtwoopt{\citeads}[3][][]{\href{http://adsabs.harvard.edu/abs/#3}%
{\def\hyper@linkstart##1##2{}%
\let\hyper@linkend\@empty\citealp[#1][#2]{#3}}}
\newcommandtwoopt{\citepads}[3][][]{\href{http://adsabs.harvard.edu/abs/#3}%
{\def\hyper@linkstart##1##2{}%
\let\hyper@linkend\@empty\citep[#1][#2]{#3}}}
\newcommandtwoopt{\citetads}[3][][]{\href{http://adsabs.harvard.edu/abs/#3}%
{\def\hyper@linkstart##1##2{}%
\let\hyper@linkend\@empty\citet[#1][#2]{#3}}}
\newcommandtwoopt{\citeyearads}[3][][]%
{\href{http://adsabs.harvard.edu/abs/#3}
{\def\hyper@linkstart##1##2{}%
\let\hyper@linkend\@empty\citeyear[#1][#2]{#3}}}
\makeatother

\begin{document}

\title{Multiwavelength study of the fast rotating supergiant high-mass X-ray binary $\igrjsqsc$\thanks{Based on observations made with ESO Telescopes at the La Silla Paranal Observatory under programme ID 077.D-0038, 077.D-0055, 077.D-0298, 077.D-0568 and 089.D-0056.}}

\author{S. Chaty\inst{1,2}
\and A. LeReun\inst{1}
\and I. Negueruela\inst{3,4} 
\and A. Coleiro \inst{5}
\and N. Castro\inst{6}
\and S. Sim\'on-D\'{\i}az\inst{7,8}  
\and J.A. Zurita Heras \inst{1}
\and P. Goldoni \inst{5}
\and A. Goldwurm \inst{5}
}
  
\authorrunning{Chaty et al.}
\titlerunning{Multiwavelength study of the sgHMXB $\igrjsqsc$}

 \offprints{S. Chaty \email{chaty@cea.fr}}
 
\institute{AIM (UMR 7158 CEA/DSM-CNRS-Universit\'e Paris Diderot), Irfu/Service d'Astrophysique, Centre de Saclay, FR-91191 Gif-sur-Yvette Cedex, France
  \and Institut Universitaire de France, 103, boulevard Saint-Michel 75005 Paris, France
  \and Departamento de F\'{\i}sica, Ingenier\'{\i}a de Sistemas y Teor\'{\i}a de la Se\~{n}al, Escuela Polit\'ecnica Superior, Universidad de Alicante, Carretera San Vicente del Raspeig s/n,  E03690, San Vicente del Raspeig, Spain
  \and Santa Cruz Institute for Particle Physics, University of California Santa Cruz, 1156 High St., Santa Cruz, CA 95064, USA
  \and APC (UMR 7164 CEA/DSM-CNRS-Universit\'e Paris Diderot), 10 rue Alice Domon et Léonie Duquet, 75013 Paris, France 
  \and Argelander Institut f\"ur Astronomie, Auf den H\"ugel 71, Bonn, 53121, Germany
  \and Instituto de Astrof\'isica de Canarias, V\'ia L\'actea s/n, E-38205 La Laguna, Santa Cruz de Tenerife, Spain
  \and Departamento de Astrof\'isica, Facultad de F\'isica y Matem\'aticas, Universidad de La Laguna, Avda. Astrof\'isico Francisco S\'anchez, s/n, E-38206 La Laguna, Santa Cruz de Tenerife, Spain
}
      
\date{\today}

\abstract {Since its launch, the X-ray and $\gamma$-ray observatory \textit{INTEGRAL} satellite has revealed a new class of high-mass X-ray binaries (HMXB) displaying fast flares and hosting supergiant companion stars. Optical and infrared (OIR) observations in a multi-wavelength context are essential to understand the nature and evolution of these newly discovered celestial objects.}
{The goal of this multiwavelength study (from ultraviolet to infrared) is to characterise the properties of $\igrjsqsc$, to confirm its HMXB nature and that it hosts a supergiant star.}
{We analysed all OIR, photometric and spectroscopic observations taken on this source, carried out at ESO facilities.}
{Using spectroscopic data, we constrained the spectral type of the companion star between B0.5 and B1\,Ib, settling the debate on the true nature of this source. We measured a high rotation velocity of $v = 320 \pm 8\:\mathrm{km}\,\mathrm{s}^{-1}$ from fitting absorption and emission lines in a stellar spectral model. We then built a spectral energy distribution from photometric observations to evaluate the origin of the different components radiating at each energy range.} 
{We finally show that, having accurately determined the spectral type of the early-B supergiant in $\igrjsqsc$, we firmly support its classification as an intermediate supergiant fast X-ray transient (SFXT).}


\maketitle 


\section{Introduction} \label{section:introduction}

High-mass X-ray binaries (HMXB) consist of a compact object orbiting a massive companion star \citep[$M \geq10\:M_{\sun}$, see e.g. reviews of][]{charles:2006,chaty:2013}. Depending on the nature of the primary star and the accretion mechanism, they can be divided into two sub-classes: HMXB hosting either a main sequence Be star (BeHMXB) or a supergiant OB star (sgHMXB). The former is characterised by a compact object revolving around a Be\,V star in a generally wide and eccentric orbit ($e\sim0.3-0.9$), where accretion usually happens in the form of outbursts close to periastron passage of the compact object. The latter consists of an OB\,I star possessing a radially outflowing wind from which the compact object on a low-to-moderate eccentricity orbit is continuously accreting.

One of the most important contributions of the INTErnational Gamma-Ray Astrophysics Laboratory satellite \cite[{\it INTEGRAL},][]{winkler:2003} is the discovery of a new population of sgHMXB, which can be divided into two subclasses \citep[further details and references in][]{chaty:2013}: intrinsically obscured sgHMXB and Supergiant Fast X-ray Transients (SFXT):

{\it i.} Multiwavelength studies of the optical and infrared (OIR) counterparts of obscured sgHMXB have revealed the presence of two envelopes. The first is a layer of X-ray absorbing material close to the compact object; the second is an absorbing envelope enshrouding the whole binary system.

{\it ii.} Hard X-ray spectra of SFXT suggest the presence of a compact object such as a neutron star or a black hole. They exhibit rapid outbursts, rising in 10 minutes and lasting typically 3\,ks, alternating with long periods of quiescence. Depending on the intensity of the X-ray luminosity and the variability factor ($\frac{L_{\mathrm{max}}}{L_{\mathrm{min}}}$), SFXT can be further separated into two subgroups: ``classical'' SFXT (low $<L_{\mathrm X}>$, high variability factor), and ``intermediate'' SFXT (high $<L_{\mathrm X}>$, low variability factor).

Among this new population of sgHMXB, $\igrjsqsc$ was discovered with {\it INTEGRAL} during its flare observed on 2004 September 6 \citep{lutovinov:2004}. From X-ray Multi-Mirror Mission ({\it XMM-Newton}) follow-up observations, \cite{zurita-heras:2004} accurately located the X-ray counterpart and proposed a near-infrared (NIR) counterpart in the 2MASS catalogue. \cite{smith:2004} identified a bright optical counterpart in the USNO catalogue, and proposed an early spectral type for the companion star. This was confirmed by \cite{negueruela:2005a}, who estimated the spectral type to B0.5\,I, and later by \cite{negueruela:2006a}, who constrained it to a luminous star in the B0-1 range\footnote{from optical observations with EMMI at ESO/NTT.}, supported by \cite{rahoui:2008a} from SED fitting results\footnote{from mid-infrared observations with VISIR at ESO/VLT.}. However, the spectral type of the companion star was a matter of debate since \cite{nespoli:2008} later proposed an O9.5\,Ia classification\footnote{from NIR observations with SOFI at ESO/NTT.}.

\cite{lutovinov:2005a} detected a pulsation (spin) period of $228 \pm 6$~s, and evaluated a strong intrinsic absorption from X-ray fitting. Both the pulsation period and high column density were confirmed by \cite{walter:2006}, who qualified this source as a ``highly absorbed supergiant transient system''. In their precursor paper, \cite{negueruela:2006a} gave the name ``SFXT'' for the members of this new class of HMXB, and suggested that $\igrjsqsc$ was one of them. Later, \cite{walter:2007} proposed that it belonged to the newly identified ``intermediate SFXT'' class. However, the SFXT classification was also debated since \cite{laparola:2010} and \cite{clark:2010} found an orbital period of $\sim 30$\,days from {\it Swift}-BAT and {\it INTEGRAL}-IBIS observations, whereas typical periods of classical SFXT were of a few days \citep[see e.g.][]{chaty:2013}. This led \cite{laparola:2010} to exclude $\igrjsqsc$ from the SFXT class, while \cite{clark:2010} supported the intermediate SFXT classification. 

This uncertainty regarding the spectral type and HMXB nature of these systems is critical because their formation and evolution both depend heavily on the nature of the companion star. This work aims to solve these issues by analysing detailed multiwavelength OIR observations both in photometry and spectroscopy. This study is based on all existing observations on this source, retrieved from the ESO archive (see Table~\ref{table:pgrm}). The paper is organised as follows. We describe in Section~\ref{section:observations} the multiwavelength observations and the data reduction process, before presenting the main results and a discussion in Section~\ref{section:results}. A summary of our conclusions is presented in Section~\ref{section:conclusion}.

\begin{table*}
\caption{OIR observation log. \label{table:pgrm}}
$$
 \centering
 \begin{tabular}{lllll} 
 \hline
\noalign{\smallskip}
Mode & Instrument & Date & Programme ID, PI & Standard stars \\
\noalign{\smallskip}
 \hline
\noalign{\smallskip}
\multirow{9}{*}{Photometry} 
& EMMI (BIMG) & 2006-Apr-14 & 077.D-0038(A), Negueruela & \\
\noalign{\smallskip}
  \cline{2-5}
\noalign{\smallskip}
& \multirow{3}{*}{SUSI2 (UBVRIZ)} & 2006-Aug-06 & \multirow{3}{*}{077.D-0298(B), Chaty} & \multirow{3}{*}{PG\,1633+099 (ABCD)} \\
& & 2006-Aug-07 & & \\
& & 2006-Sep-04 & & \\ 
\noalign{\smallskip}
\cline{2-5}
\noalign{\smallskip}
& \multirow{5}{*}{SOFI (JHK$_{\mathrm{s}}$)} & 2006-Jul-19 & \multirow{5}{*}{077.D-0298(A), Chaty} & \multirow{5}{*}{SJ\,9105, 9160, 9170, 9172, 9185, 9187} \\
& & 2006-Aug-06 & & \\
& & 2006-Aug-07 & & \\
& & 2006-Aug-08 & & \\
& & 2006-Sep-04 & & \\ 
\noalign{\smallskip}
\hline
\multirow{9}{*}{Spectroscopy} 
& FORS1 & 2006-Apr-24 & 077.D-0055(A), Negueruela & \\
\noalign{\smallskip}
  \cline{2-5}
\noalign{\smallskip}
& EMMI (RILD) & 2006-Aug-07 & 077.D-0298(C), Chaty & LTT\,7379 (G0\,V) \\
\noalign{\smallskip}
  \cline{2-5}
\noalign{\smallskip}
& \multirow{6}{*}{SOFI (GBF/GRF)} 
  & 2006-Jul-18 & \multirow{4}{*}{077.D-0298(A), Chaty} & Hip\,078791 (G2\,V) \\ 
& & 2006-Jul-19 & & Hip\,078791 (G2\,V) \\ 
& & 2006-Aug-08 & & Hip\,084419 (= HD\,155755, G2\,V) \\ 
& & 2006-Sep-03 & & Hip\,081746 (= HD\,150248, G3\,V)\\ 
\noalign{\smallskip}
  \cline{3-5}
\noalign{\smallskip}
& & 2006-Jul-14 & \multirow{2}{*}{077.D-0568(A), Mennickent} & \\
& & 2006-Jul-26 & & \\
\noalign{\smallskip}
\cline{2-5}
\noalign{\smallskip}
& X-shooter & 2012-Aug-07 & 089.D-0056(A), Goldwurm & \\
\noalign{\smallskip}
\hline
\noalign{\smallskip}
\end{tabular}
$$
\end{table*}  


\section{Observations} \label{section:observations}


   \subsection{Photometry}

Images were obtained at the European Southern Observatory (ESO) at the focus of the New Technology Telescope (NTT-3.5\,m, La Silla, Chile), with three instruments: Extraordinaire Multi-Mode Instrument \cite[EMMI][]{dekker:1986} and Superb Seeing Imager \cite[SUSI2][]{dodorico:1998} in optical, and Son OF Isaac \cite[SOFI][]{moorwood:1998} in NIR (dates listed in Table~\ref{table:photdate}). Images were acquired in large field mode covering on the sky $6\farcm2 \times 6\farcm2$ (EMMI), $5\farcm5\times 5\farcm5$ (SUSI2), and $4\farcm9\times 4\farcm9$ (SOFI).

Data reduction was standard, using the Image Reduction and Analysis Facility ({\it IRAF}) suite \citep{tody:1986,tody:1993}. Bias and flat corrections were applied in both wavelength domains. NIR images were taken at nine positions, enabling us to evaluate and remove the sky brightness. Images of the standard stars PG\,1633+099(ABCD) in the optical and SJ\,9105, 9160, 9170, 9172, 9185, and 9187 in the NIR enabled us to perform photometry using the {\it IRAF.daophot} package. 
We calculated the apparent magnitudes by subtracting the zero-point and extinction coefficient multiplied by the airmass at the time of the observations from the instrumental magnitudes, using the standard formula. Extinction coefficients were taken from the calibration plan of La Silla Observatory.

We present in Tables~\ref{table:mag-opt} and \ref{table:mag-nir} the results of our photometry, which represent the first whole set of OIR magnitudes published on this source\footnote{apart from the USNO and 2MASS catalogues.}.

\begin{table*}
\caption{Photometric observations with their cumulated exposure time.\label{table:photdate}}
$$
\centering
 \begin{tabular}{llc c c c c c c c c c cll} 
 \hline
\noalign{\smallskip}
 \multirow{2}{*}{Date} &  \multirow{2}{*}{Instrument} & \multicolumn{9}{ c }{Filter} & \\
\noalign{\smallskip}
 \cline{3-11}
\noalign{\smallskip}
  &  & U & B & V & R & I & Z & J & H & K$_{\mathrm{s}}$ & \\
\noalign{\smallskip}
 \hline 
\noalign{\smallskip}
  2006-Apr-14& EMMI  & -   & 30\,s & -   & 1\,s  & -   & - & - & - & - &\\ 
 \noalign{\smallskip}
\hline 
\noalign{\smallskip}
  2006-Aug-07 & SUSI2 & 60\,s & 60\,s & 60\,s & 60\,s & 60\,s & 60\,s & - & - & - &\\ 
\noalign{\smallskip}
 \hline
\noalign{\smallskip}
  2006-Sep-04 & SUSI2 & 60\,s & 60\,s & 60\,s & 60\,s & 20\,s & - & - & - & - &\\ 
\noalign{\smallskip}
 \hline
\noalign{\smallskip}
  2006-Jul-19 & SOFI  & -   & -   & -   & -   & -   & - & $9 \times 36$\,s & $9 \times 20$\,s & $9 \times 18$\,s &  \\
\noalign{\smallskip}
 \hline
\noalign{\smallskip}
  2006-Aug-06 & SOFI  & -   & -   & -   & -   & -   & - & $9 \times 18$\,s & - & - & \\ 
\noalign{\smallskip}
 \hline
\noalign{\smallskip}
  2006-Aug-07 & SOFI  & -   & -   & -   &  -  & -   & - & - & $9 \times 20$\,s & $9 \times 10.8$\,s &  \\ 
\noalign{\smallskip}
 \hline
\noalign{\smallskip}
  2006-Aug-08 & SOFI  & -   & -   & -   & -   & -   & - & $9 \times 36$\,s & $9 \times 13.5$\,s & $9 \times 10.8$\,s &  \\ 
\noalign{\smallskip}
 \hline
\noalign{\smallskip}
  2006-Sep-04 & SOFI  & -   & -   & -   & -   & -   & - & $9 \times 10.6$\,s & $9 \times 10.6$\,s & $9 \times 7.1$\,s &  \\ 
\noalign{\smallskip}
 \hline
\end{tabular}
$$
 \end{table*}

\begin{table*}
\caption{Optical photometry of $\igrjsqsc$.\label{table:mag-opt}}
\centering
 \begin{tabular}{llcccccc} 
 \hline
 \noalign{\smallskip}
 Date & Instrument & U  &  B  &  V  &  R  &  I  &  Z  \\
 \hline
 \noalign{\smallskip}
 2006-Apr-14 & EMMI  &                & $17.00\pm0.01$ &                & $12.71\pm0.01$ &                &                \\ 
            &airmass&                & 1.230          &                & 1.250          &                &                \\ 
\noalign{\smallskip}
 \hline
\noalign{\smallskip}
 2006-Aug-07 & SUSI2 & 17.11$\pm$0.12 & 16.86$\pm$0.01 & 14.64$\pm$0.01 & 13.33$\pm$0.01 & 12.34$\pm$0.01 & 13.23$\pm$0.01 \\ 
            &airmass& 1.281          & 1.287          & 1.293          & 1.299          & 1.304          & 1.310          \\ 
\noalign{\smallskip}
 \hline
\noalign{\smallskip}
 2006-Sep-04 & SUSI2 & 17.15$\pm$0.01 & 16.85$\pm$0.01 & 14.62$\pm$0.01 & 13.55$\pm$0.01 & 12.45$\pm$0.01 & \\
            &airmass& 1.168          & 1.172          & 1.176          & 1.189          & 1.192          &               \\ 
\noalign{\smallskip}
 \hline
\noalign{\smallskip}
 \multicolumn{2}{l}{USNO\,0448-0520455} & & 15.20 & & R1=12.70 & 11.74 & \\  
                                      & & &       & & R2=13.02 &       & \\
\noalign{\smallskip}
 \hline
\end{tabular} 
 \end{table*}

\begin{table*}
\caption{NIR photometry of $\igrjsqsc$.\label{table:mag-nir}}
\centering
 \begin{tabular}{llccc} 
 \hline
 \noalign{\smallskip}
 Date & Instrument & J  &  H  &  K$_{\mathrm{s}}$ \\
 \noalign{\smallskip}
 \hline
\noalign{\smallskip}
 2006-Jul-19 & SOFI  & 10.53$\pm$0.03 & 10.03$\pm$0.06  & 9.81$\pm$0.09 \\  
            &airmass& 1.049          & 1.044          & 1.042         \\
 \noalign{\smallskip}
 \hline
\noalign{\smallskip}
 2006-Aug-06 & SOFI  & 10.51$\pm$0.05 & & \\ 
            &airmass& 1.052          & & \\
\noalign{\smallskip}
 \hline
\noalign{\smallskip}
 2006-Aug-07 & SOFI  &                & 10.03$\pm$0.06 & 9.81$\pm$0.06 \\
            &airmass&                & 1.043          & 1.040         \\ 
\noalign{\smallskip}
 \hline
\noalign{\smallskip}
 2006-Aug-08 & SOFI  & 10.56$\pm$0.05 & 10.02$\pm$0.05 & 9.90$\pm$0.06 \\ 
            &airmass& 1.254          & 1.299          & 1.377         \\ 
 \hline
 \noalign{\smallskip}
 2006-Sep-04 & SOFI  & 10.53$\pm$0.05 & 10.01$\pm$0.06 & 9.83$\pm$0.07 \\  
            &airmass& 1.296          & 1.367          & 1.447         \\ 
\noalign{\smallskip}
 \hline
\noalign{\smallskip}
 \multicolumn{2}{l}{2MASS\,J16463526-4507045} 
                    &10.537$\pm$0.023&10.079$\pm$0.025&9.843$\pm$0.023\\
\noalign{\smallskip}
 \hline
\end{tabular}
 \end{table*}


   \subsection{Spectroscopy}

      \subsubsection{Optical (FORS1+EMMI) and NIR (SOFI) observations}

Optical long slit spectroscopy was performed in service mode on 2006 April 24 at the Very Large Telescope (VLT, Paranal, Chile) Unit Telescope Kueyen (UT2), equipped with the spectro-imager FOcal Reducer and low dispersion Spectrograph (FORS1), fitted with a thinned Tektronix 24\,$\mu$m $2048\times2048$ pixel CCD, with an exposure time of 1670~s. We used the holographic grism 1200B+97, providing a nominal dispersion of 0.06\,nm/pixel over the $372 - 486$\,nm range, and a slit $1\farcs3$ in width, which led to a spectral resolution of R\,$= \frac{\lambda}{\Delta \lambda} = 1420$. A calibration spectrum was obtained with the same setup. Additional optical spectra ($6\times300$~s) were acquired with EMMI on ESO/NTT in Red Imaging and Low Dispersion Spectrum (RILD) mode, in the red arm through an $1\arcsec$ slit on 2006 August 7 (spectral resolution R\,$= 613$).

NIR long slit spectroscopy was also obtained at ESO/NTT using SOFI low resolution grisms, between 2006 July 14 and September 3: 36 spectra $\times$ 60~s exposure time taken with Grism Blue Filter (GBF, $\Delta \lambda = 0.95-1.64 \microns$, R\,$\sim 930$, dispersion\,=\,0.696\,nm/pixel), and 36 spectra $\times$ 60~s exposure time taken with Grism Red Filter (GRF, $\Delta \lambda = 1.53-2.52 \microns$, R\,$\sim 980$, dispersion\,=\,1.022\,nm/pixel), both through a $0\farcs6$ slit. 

\begin{figure}
\centering
\includegraphics[width=1.1\textwidth]{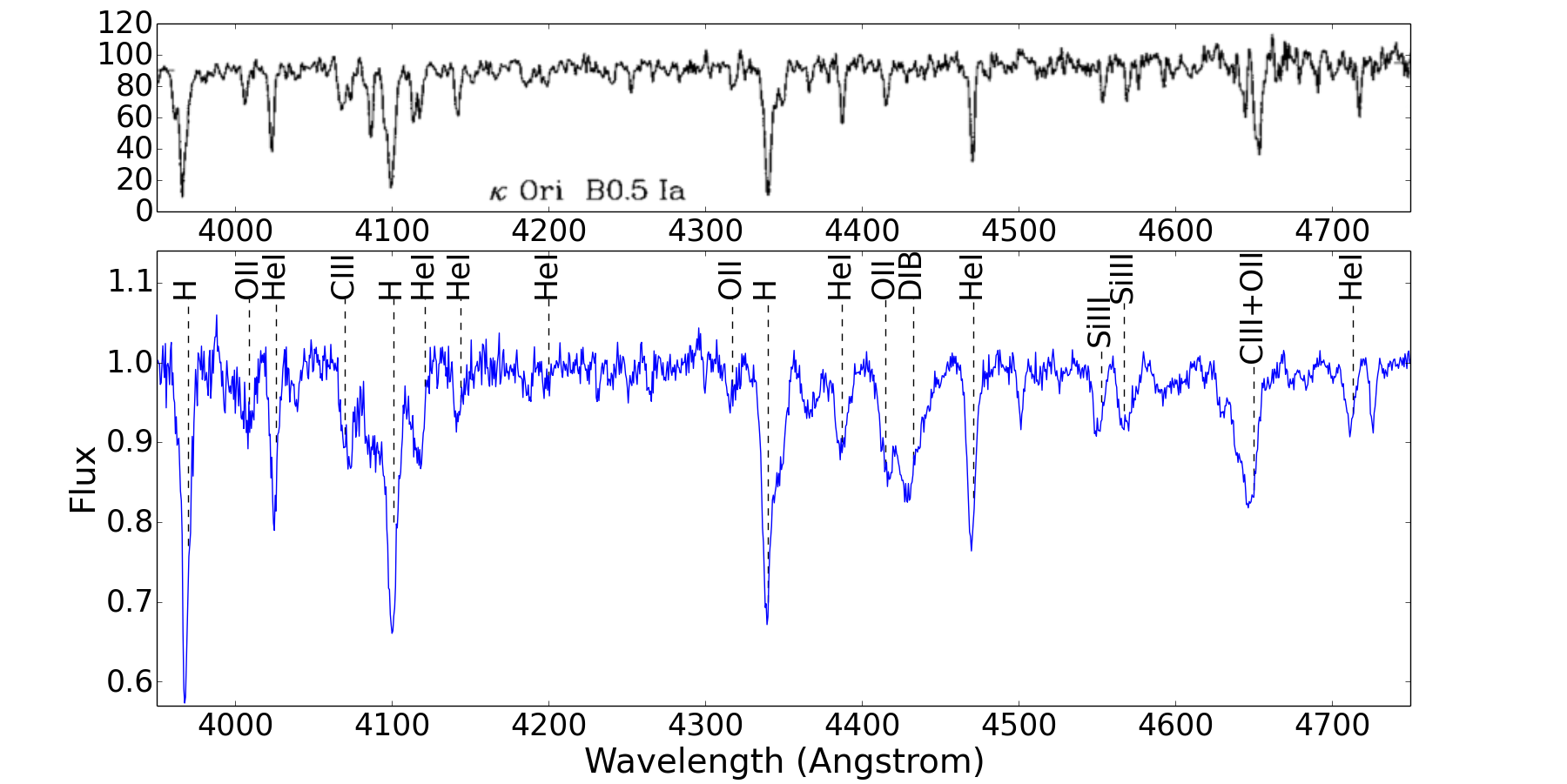}
\caption{FORS1 spectrum of $\igrjsqsc$ from 395 to 475\,nm (flux axis is in arbitrary units). For comparison, a spectrum of the B0.5\,I star {\it k} Ori is shown in the upper panel \citep{walborn:1990}.
\label{fig:FORS1optID}}
\end{figure}

\begin{figure}
\centering
\includegraphics[width=1.1\textwidth]{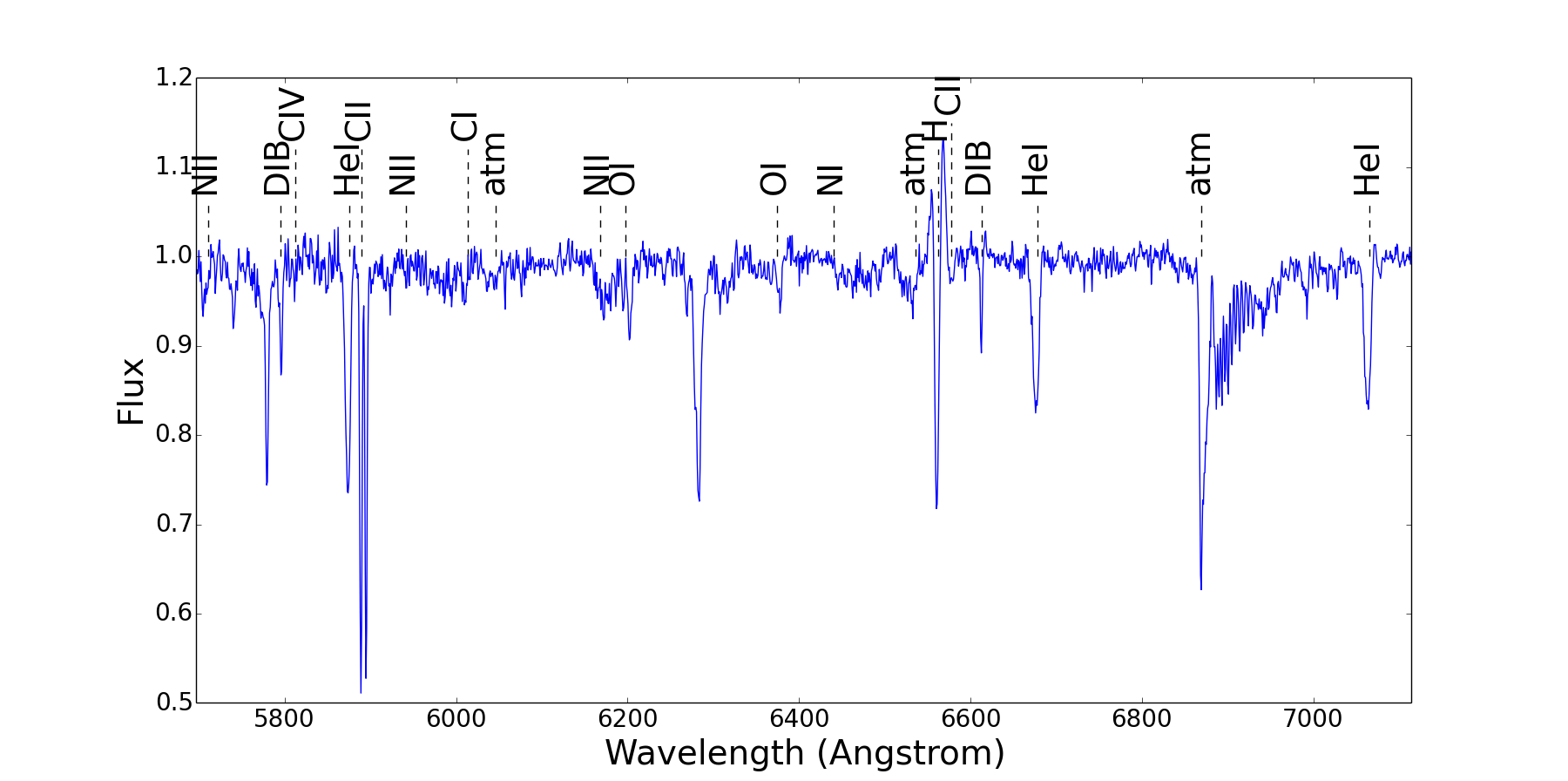}
\caption{EMMI spectrum of $\igrjsqsc$ from 570 to 710\,nm (flux axis is in arbitrary units). 
\label{fig:EMMIspec}}
\end{figure}

\begin{figure}
\centering
\includegraphics[width=1.0\textwidth]{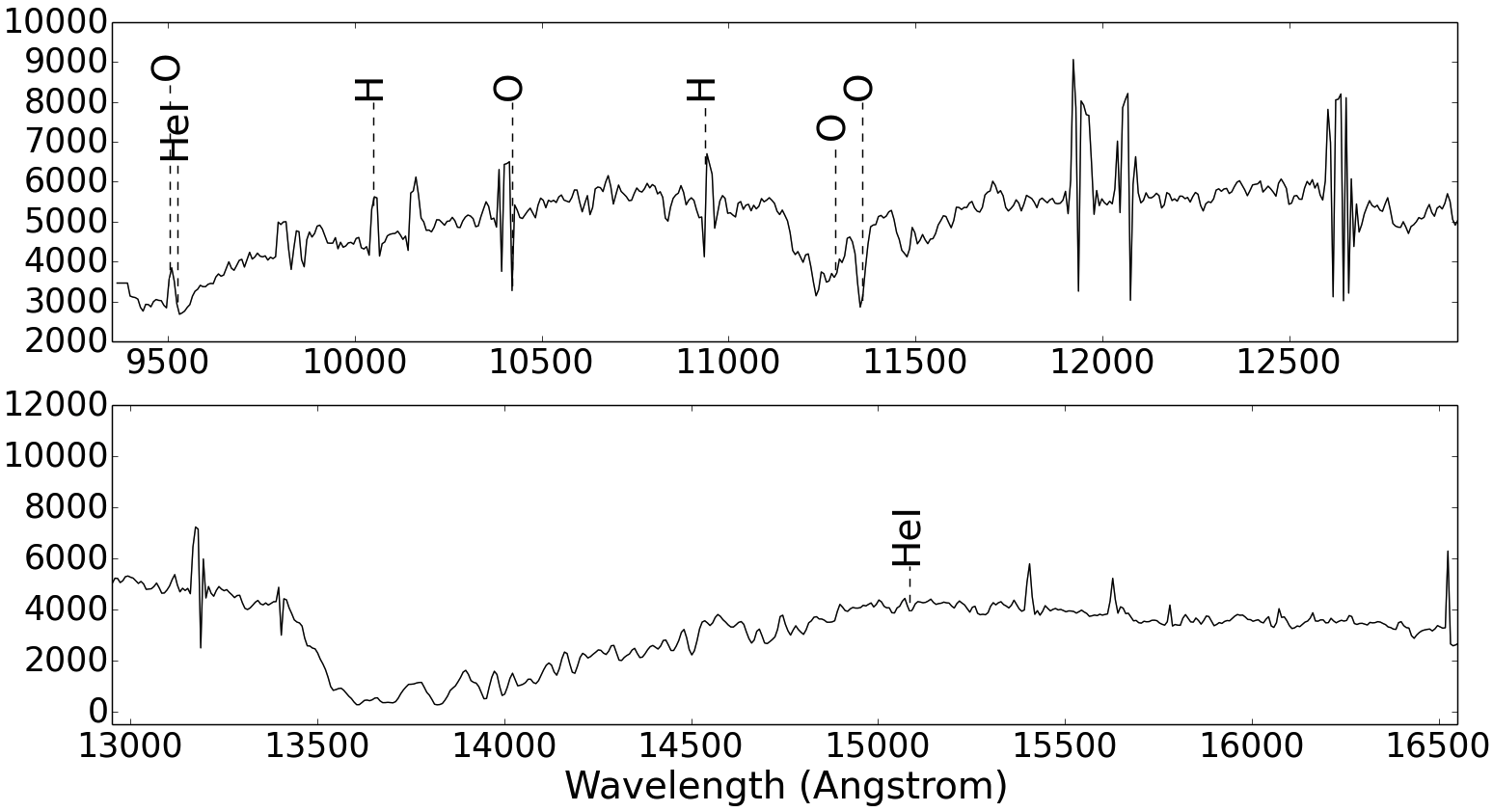}
\caption{SOFI/GBF spectrum of $\igrjsqsc$ from 950 to 1650\,nm (flux axis is in arbitrary units). Features at $\sim$ 1190, 1210, and 1265\,nm\ are due to strong atmosphere absorption in the J band.
\label{fig:SOFIgbfspec}}
\end{figure}

\begin{figure}
\centering
\includegraphics[width=1.0\textwidth]{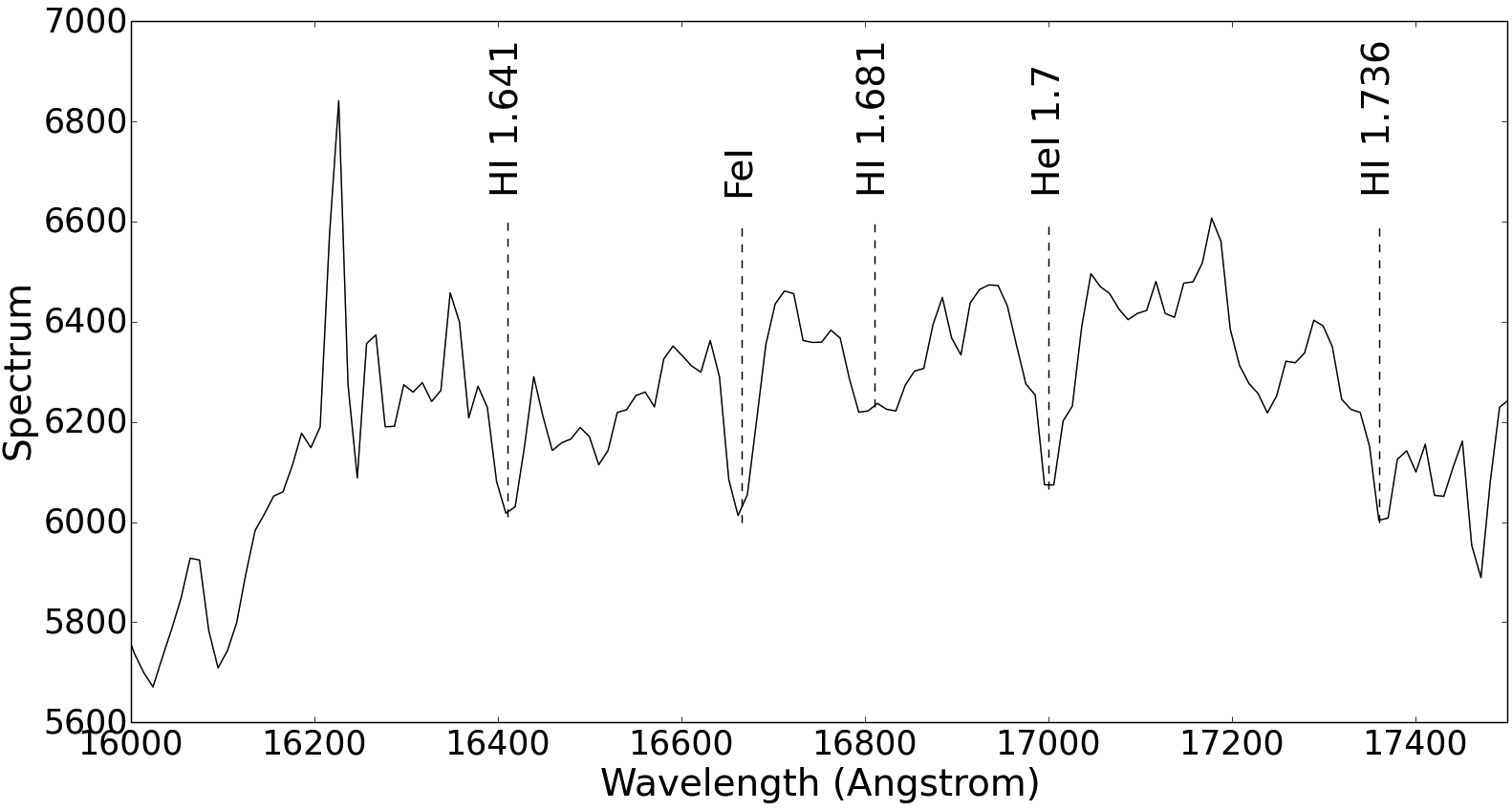}
\caption{SOFI/GRF (H part) spectrum of $\igrjsqsc$ from 1600 to 1750\,nm (flux axis is in arbitrary units). The feature at $\sim 1625$\,nm is probably due to a bad background correction at the edge of the grism.
\label{fig:SOFIgrfspec_H}}
\end{figure}

\begin{figure}
\centering
\includegraphics[width=1.0\textwidth]{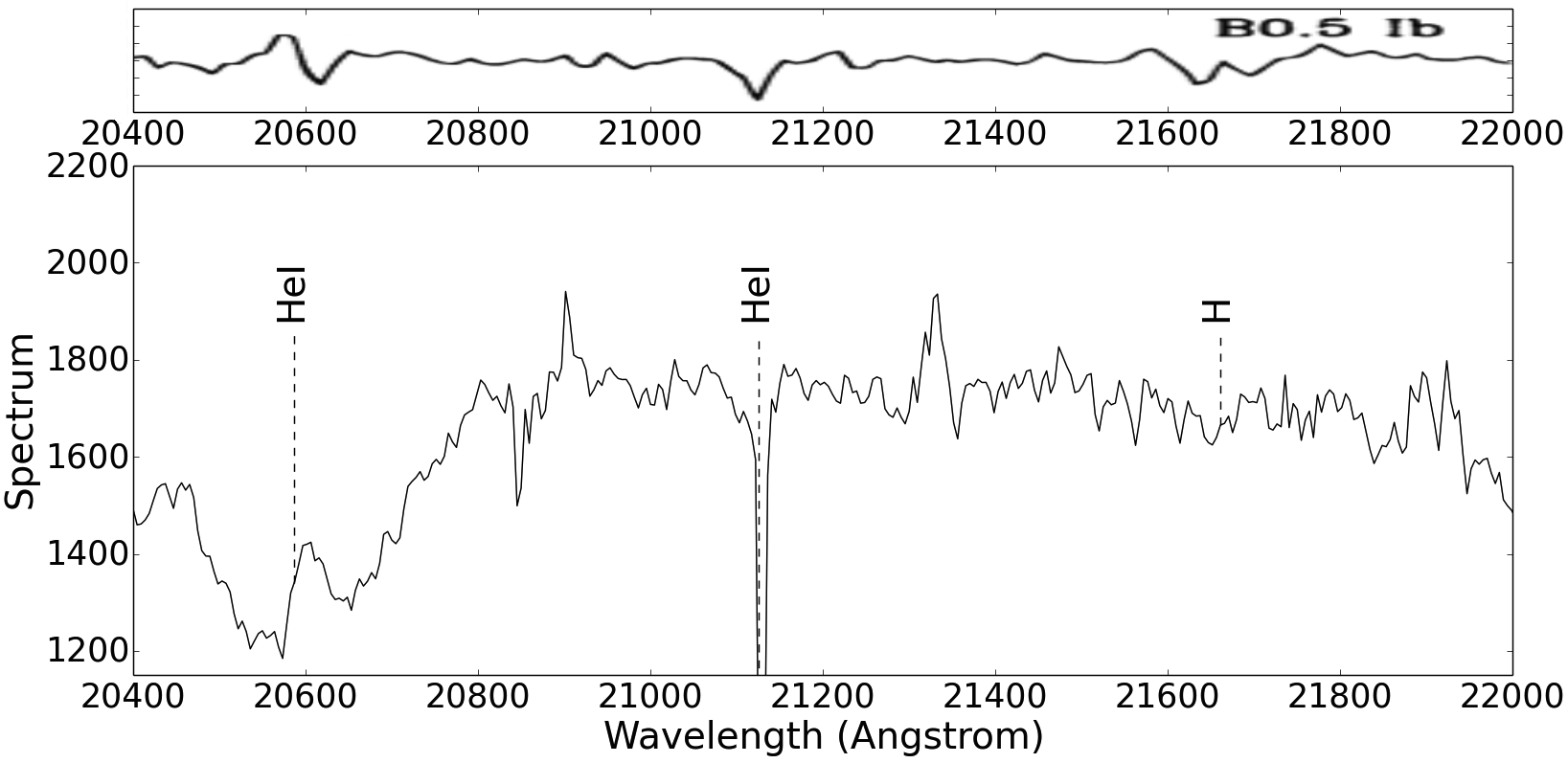}
\caption{SOFI/GRF ($K_{\mathrm{s}}$ part) spectrum of $\igrjsqsc$ from 2040 to 2200\,nm. A typical B0.5\,Ib spectrum is indicated at the top of the figure (flux axis is in arbitrary units for both spectra).
\label{fig:SOFIgrfspec_Ks}}
\end{figure}

The FORS1 spectrum was reduced with the {\em Starlink} packages {\sc ccdpack} \citep{draper:2000} and {\sc Figaro} \citep{shortridge:1997}. 
EMMI and SOFI spectra were reduced using {\it IRAF} in a standard way. We first performed bias and flat correction, and removed cosmic rays using {\it IRAF.lacos} task. We then applied wavelength calibration before subtracting the telluric sky lines, and performed spectra extraction with {\it IRAF.noao.twodspec} package. Finally, we combined all the spectra for each instrument together to increase the signal-to-noise ratio. FORS1, EMMI, and SOFI spectra are shown in Figures\,\ref{fig:FORS1optID}, \ref{fig:EMMIspec}, \ref{fig:SOFIgbfspec}, \ref{fig:SOFIgrfspec_H}, and \ref{fig:SOFIgrfspec_Ks}; the positions of identified spectral features are indicated by dashed lines. Many hydrogen and helium lines are seen in absorption and some of them present a shape typical of self-absorbed lines. Bright metal lines are also identified in some spectra.

      \subsubsection{Ultraviolet to NIR X-shooter observations}

The X-shooter\footnote{X-shooter is the first second-generation instrument installed in 2008 at the ESO VLT.} instrument is the successor spectrograph of FORS1, mounted at the UT2 cassegrain focus \citep{vernet:2011}. Thanks to its three arms --UV-blue (UVB), visible (VIS) and near-IR (NIR)--, the main characteristic of this medium-resolution spectrograph is its ability to cover, in a single observation, a wide wavelength range from 300 to 2480\,nm. In July 2012 we obtained guaranteed time on X-shooter to observe a small sample of bright {\it INTEGRAL} counterparts in order to better constrain their nature. Preliminary results on three sources, including $\igrjsqsc$, have been reported in \cite{goldoni:2012}. 

All the spectra were taken with narrow slits: slit width $0\farcs5$ and sampling 3.5\,pix/FWHM (full width half maximum) for the UVB arm, slit width $0\farcs7$ and sampling 4.8\,pix/FWHM for the VIS arm, and slit width $0\farcs6$ and sampling 2.9\,pix/FWHM for the NIR arm. As described in \cite{goldoni:2012}, we also took four different pointings of 300\,s each in the echelle slit-nod mode with an offset of $5\asec$ along slits between pointings in a standard ABBA sequence. To avoid saturation, the exposures on the VIS and NIR arms were split into shorter integrations. In addition, for calibration purposes each source was briefly observed with a wide ($5\farcs0$) slit in order to estimate the slit losses. A telluric A0V star was observed before each source, and a flux standard was observed at the beginning of the night.

The spectra were retrieved from the ESO archive, and processed using the X-shooter pipeline \citep{modigliani:2010}\footnote{These so-called Phase 3 data are reduced using slit-nod related recipes, including flat and bias correction, and wave and flux calibration.}. We clearly detected the source in each of the three arms. All X-shooter spectra are given in nm (Figs.~\ref{fig:xshuvbspec}--\ref{fig:xshnirspec}). 

All the identified lines are listed in the Appendix (Tables~\ref{table:forsline}  to \ref{table:xtabnir}).

\begin{figure}
\centering
\includegraphics[width=1.0\textwidth]{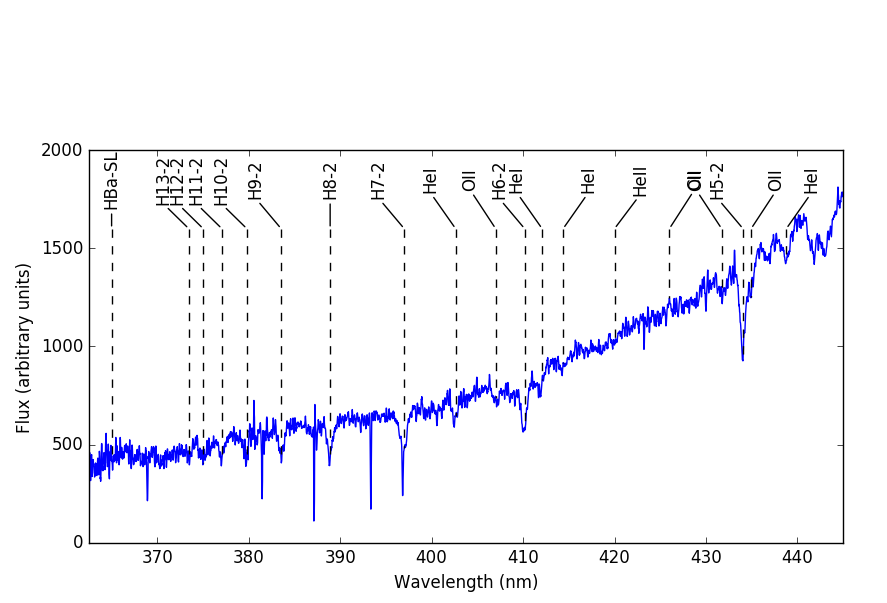}
\includegraphics[width=1.0\textwidth]{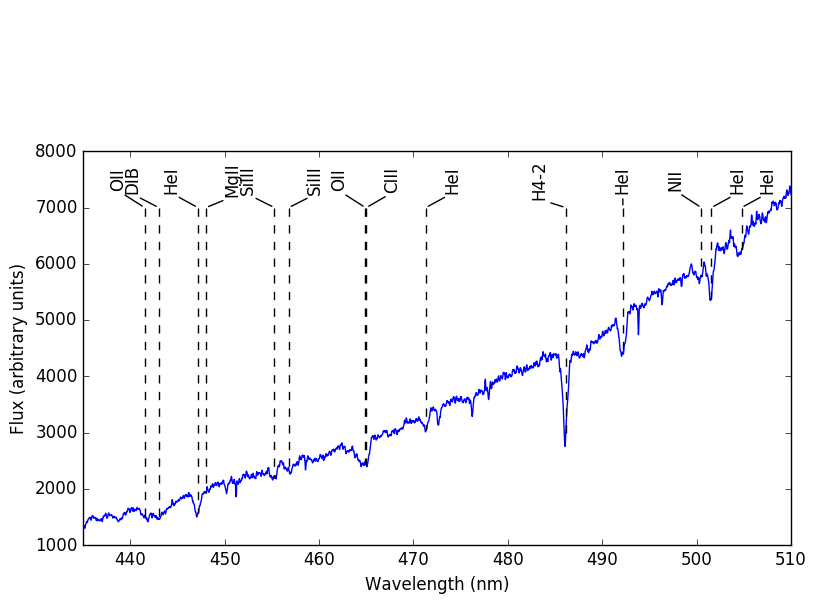}
\caption{X-shooter/UVB spectrum of $\igrjsqsc$ from 360 to 510\,nm. DIB stands for diffuse interstellar band.} 
\label{fig:xshuvbspec}
\end{figure}  

\begin{figure}
\centering
\includegraphics[width=1.0\textwidth]{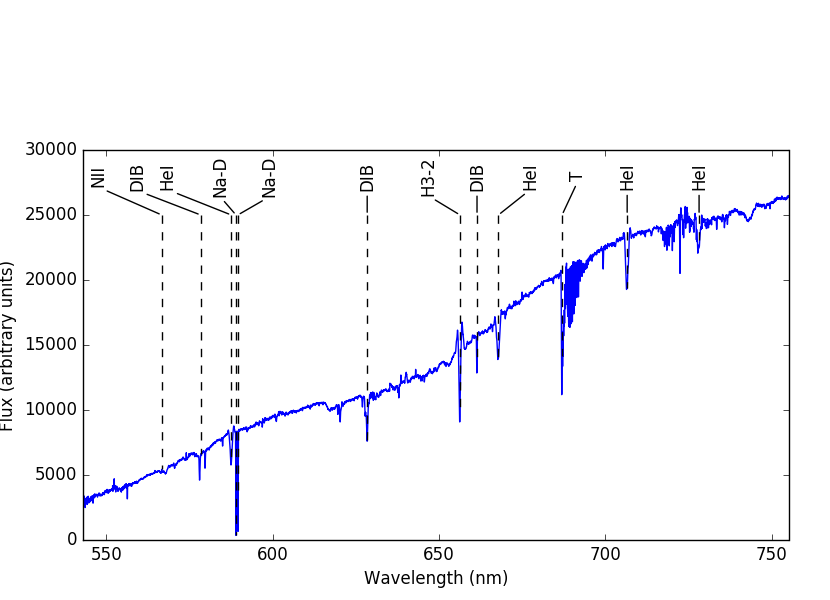}
\includegraphics[width=1.0\textwidth]{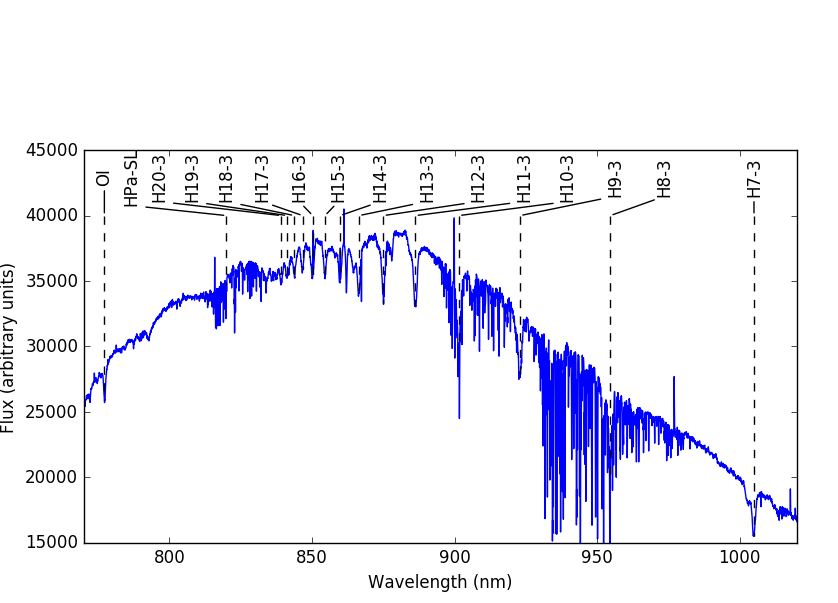}
\caption{X-shooter/VIS spectrum of $\igrjsqsc$ from 540 to 1020\,nm. DIB stands for diffuse interstellar band, and T telluric lines.} 
\label{fig:xshvisspec}
\end{figure}

\begin{figure}
\centering
\includegraphics[width=1.0\textwidth]{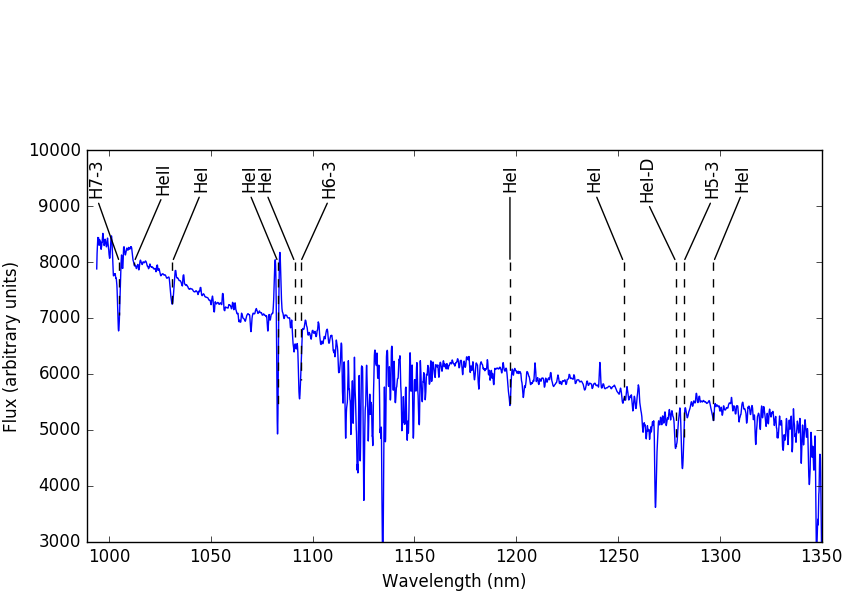}
\includegraphics[width=1.0\textwidth]{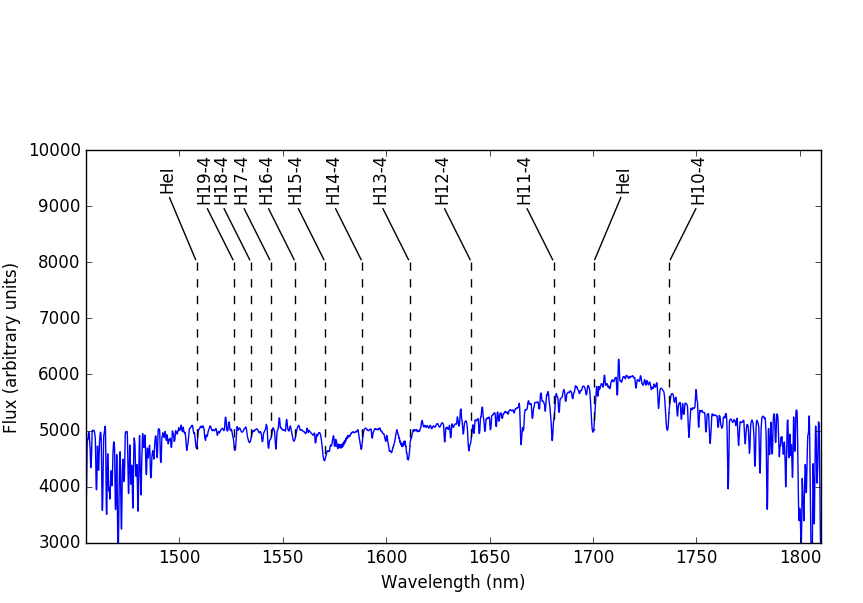}
\includegraphics[width=1.0\textwidth]{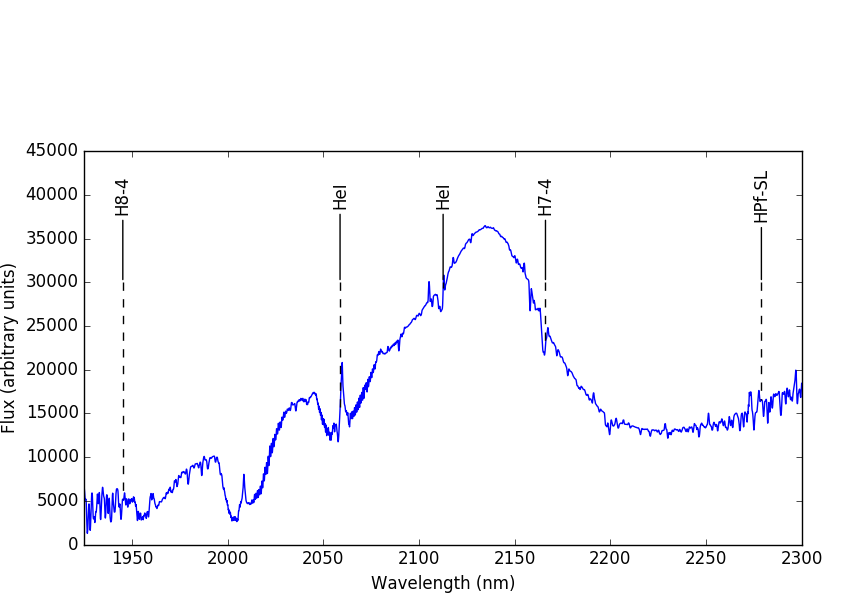}
\caption{X-shooter/NIR spectrum of $\igrjsqsc$ from 990 to 2300\,nm.}
\label{fig:xshnirspec}
\end{figure}


\section{Results and discussion} \label{section:results}


   \subsection{Spectral type determination}


      \subsubsection{FORS1 and X-shooter optical observations}

We first determined the spectral type in the optical domain, comparing the main lines identified in our FORS1 and X-shooter/UVB spectra (Fig.~\ref{fig:FORS1optID} and \ref{fig:xshuvbspec} respectively) with the optical stellar spectral atlas from \cite{walborn:1990}. We note that the \ion{He}{i} and \ion{Si}{iii} lines, clearly detected in both spectra, are more intense than the \ion{He}{ii} and \ion{Si}{iv} lines, which is consistent with an early-B type. Then, the H$\gamma$~434.1\,nm line, present in both spectra, appears blended with a prominent \ion{O}{ii} line (see Fig.~\ref{fig:xshuvbspec}), which clearly exhibits a structure typical of spectral types close to B1\,Ib. 

In addition, by examining the X-shooter/VIS spectrum reported in Fig.~\ref{fig:xshvisspec}, we realised that the Paschen line P16, detected at 850.0\,nm, must be blended with the \ion{C}{iii}~850.2\,nm line because its intensity is higher than that of other Paschen lines at longer wavelengths (i.e. lower members of the same series). This detection of carbon in the spectrum suggests a spectral type earlier than B1\,I \citep[see e.g.][]{negueruela:2010a}. 
Furthermore, the presence of only a very weak \ion{He}{ii} line at $\sim 420$\,nm (see X-shooter/UVB spectra)\footnote{We identify the line close to $\sim 420$\,nm as \ion{He}{ii} instead of \ion{N}{ii}, based on the absence of other nitrogen lines in our spectrum \cite[see e.g. the spectrum of BD~$+36\adeg 4063$ in][]{walborn:2000}} shows that the spectral type cannot be earlier than B0\,I \cite[see e.g.][]{walborn:1990}.

Finally, we observed that the optical spectrum of $\igrjsqsc$ is very similar to the typical B0.5\,I spectrum indicated at the top of Fig.~\ref{fig:FORS1optID}. Thus from all of the above we can constrain the spectral type to a luminous supergiant star of spectral type between B0.5\,I and B1\,I. This result is in agreement with the preliminary X-shooter spectra of this source reported in \cite{goldoni:2012}.


      \subsubsection{SOFI and X-shooter NIR observations}

We then completed our determination of the spectral type of the companion star by comparing our NIR spectra with the atlas of stellar spectral types in NIR reported in \cite{hanson:1996}. We identified the main features in the wavelength range from $2.04\microns$ to $2.2\microns$ (see SOFI/GRF and X-shooter/NIR spectra in Fig.~\ref{fig:SOFIgrfspec_Ks} and \ref{fig:xshnirspec}). \ion{He}{i}~$2112.5$\,nm and \ion{H}{i}~$2166.1$\,nm are seen in absorption, which is typical of late-O and early-B stars. On the other hand, \ion{He}{i}~$2058.9$\,nm is in emission, which is incompatible with the O9.5 type proposed by \cite{nespoli:2008}. Instead, we notice that the NIR spectrum of $\igrjsqsc$ is very similar to the typical low-resolution B0.5\,Ib spectrum indicated at the top of Fig.~\ref{fig:SOFIgrfspec_Ks}.

We thus conclude that the companion star of $\igrjsqsc$ exhibits a spectral type between B0.5\,Ib and B1\,Ib, which is in agreement with our determination above of the spectral type based on optical spectra. Finally, this result is consistent with the preliminary identification and with the classification as an SFXT first suggested by \cite{negueruela:2006a}.

 
   \subsection{Rotation and expansion velocity}

We determine the stellar rotation velocity using the classical formula of the Doppler-Fizeau effect, applied by measuring the FWHM of several H and He lines, and making the appropriate correction for the differing central wavelengths of each line, following \cite{steele:1999}:

 \begin{equation}
 v = v_{r} \times \sin i \sim c \times \frac{FWHM}{2 \lambda}.
 \label{eq:dopform}
 \end{equation}
Here $v_r$ is the radial velocity, i the inclination angle, $c$ the velocity of light, and $\lambda$ the wavelength.

The results are listed in Table ~\ref{table:speedV}. Despite a large dispersion, we calculated from this set of values a median rotation velocity of $v = 430\:\mathrm{km}\,\mathrm{s}^{-1}$. This value is extremely high, considering typical stellar rotation velocities in sgHMXB between $v = 50$ and $v = 150\:\mathrm{km}\,\mathrm{s}^{-1}$ \citep{liu:2006}. While \cite{goldoni:2012} have already reported a rotation velocity greater than $v = 200\:\mathrm{km}\,\mathrm{s}^{-1}$, the high inclination they invoked cannot be the only reason for this unusually high value.

We report in Section \ref{modelisation} below that more elaborate modelling using X-shooter spectra give a velocity $v  = 320 \pm 8\:\mathrm{km}\,\mathrm{s}^{-1}$. Our observations thus suggest that the companion star of $\igrjsqsc$ is a luminous supergiant star of spectral type B0.5--B1\,Ib with very high rotation velocity. Therefore, the star must have a small radius, typically of $\sim 15 R_{\sun}$, in order not to be disrupted by such a high stellar rotation velocity.

In addition, we clearly detect additional emission in the wings of H$\alpha$ (\ion{H}{i}\,$656.2$\,nm) both in EMMI (Fig.~\ref{fig:EMMIspec}) and X-shooter/VIS (Fig.~\ref{fig:xshvisspec}) spectra, and we also identify a P-Cygni profile at \ion{H}{i}\,$1093.8$\,nm in both SOFI/GBF (Fig.~\ref{fig:SOFIgbfspec}) and X-shooter/NIR (Fig.~\ref{fig:xshnirspec}) spectra. These features are probably due to circumstellar material (disk-like?) in a similar way to IGR~J16318-4848 \citep{chaty:2012a}, giving an estimate of the expansion velocity of the envelope: $v  = 160-180\:\mathrm{km}\,\mathrm{s}^{-1}$.

\begin{table}
\caption{Rotation velocity measurement obtained by fitting spectral line width (the associated error is of the order of 10\%). \label{table:speedV}}
$$
 \centering
 \begin{tabular}{l c c l} 
 \hline
Line (nm)   & FWHM (nm) & $v$ ($\mathrm{km}\,\mathrm{s}^{-1}$) & Instrument\\ 
 \hline
H\,397.0    & 1.06 & 400.0 & FORS1 \\
 \hline
He\,402.7   & 1.34 & 498.1 & FORS1 \\
 \hline
H\,410.2    & 1.42 & 518.8 & FORS1 \\ 
\hline
He\,412.1   & 1.44 & 523.4 & FORS1 \\
\hline
He\,438.8   & 1.64 & 560.5 & FORS1 \\
 \hline
He\,447.2   & 1.39 & 465.5 & FORS1 \\
 \hline
He\,471.3   & 1.55 & 494.2 & FORS1 \\
 \hline
H\,1005.1   & 2.86 & 426.8 & SOFI (GBF) \\
 \hline
He\,1508.3  & 3.26 & 324.2 & SOFI (GBF) \\
 \hline
H\,1641.0   & 6.25 & 571.3 & SOFI (GRF) \\
 \hline
He\,2058.9  & 8.20 & 597.4 & SOFI (GRF) \\ 
 \hline 
He\,402.0   & 1.15 & 430.0 & X-shooter/UVB \\
 \hline
H\,410.2    & 1.16 & 424.2 & X-shooter/UVB\\
 \hline
H\,434.0    & 1.38 & 477.3 & X-shooter/UVB \\
 \hline
He\,438.8   & 1.94 & 666.0 & X-shooter/UVB \\
 \hline
He\,447.0   & 1.32 & 444.3 & X-shooter/UVB \\
 \hline
He\,471.0   & 1.27 & 403.8 & X-shooter/UVB \\
 \hline
H\,486.0    & 1.36 & 420.7 & X-shooter/UVB \\
 \hline
He\,492.0   & 1.32 & 403.4 & X-shooter/UVB \\
 \hline
He\,504.0   & 1.44 & 427.9 & X-shooter/UVB \\
 \hline
H\,656.0    & 1.34 & 307.3 & X-shooter/VIS \\
 \hline
He\,667.9   & 1.84 & 413.5 & X-shooter/VIS \\
 \hline
He\,2112.6  & 6.77 & 276.9 & X-shooter/NIR \\
 \hline
\end{tabular}
$$
\end{table}

   \subsection{Stellar spectrum modelling} \label{modelisation}

We applied the {\sc iacob-broad} Interactive Data Language (IDL) tool described in \cite{simon-diaz:2014} to estimate the projected rotational velocity of the star and the amount of macroturbulent broadening ($\Theta_{{\rm RT}}$) affecting the line profiles. Unfortunately, metallic lines are very broad and often blended. The determination was thus carried out with the \ion{He}{i}~492.2\,nm line (X-shooter/UVB spectrum in Fig.~\ref{fig:xshuvbspec}). We derived the value $v = v_{r} \times \sin i = 320 \pm 8\:\mathrm{km}\,\mathrm{s}^{-1}$. This value for the rotational velocity is much higher than that found for objects of similar spectral type with the same method \citep{simon-diaz:2014}. For such high projected rotational velocity, the value of $\Theta_{\rm RT}$ is negligible and was thus taken as zero.

\begin{figure*}
\centering
\includegraphics[width=0.75\textwidth,angle=+90]{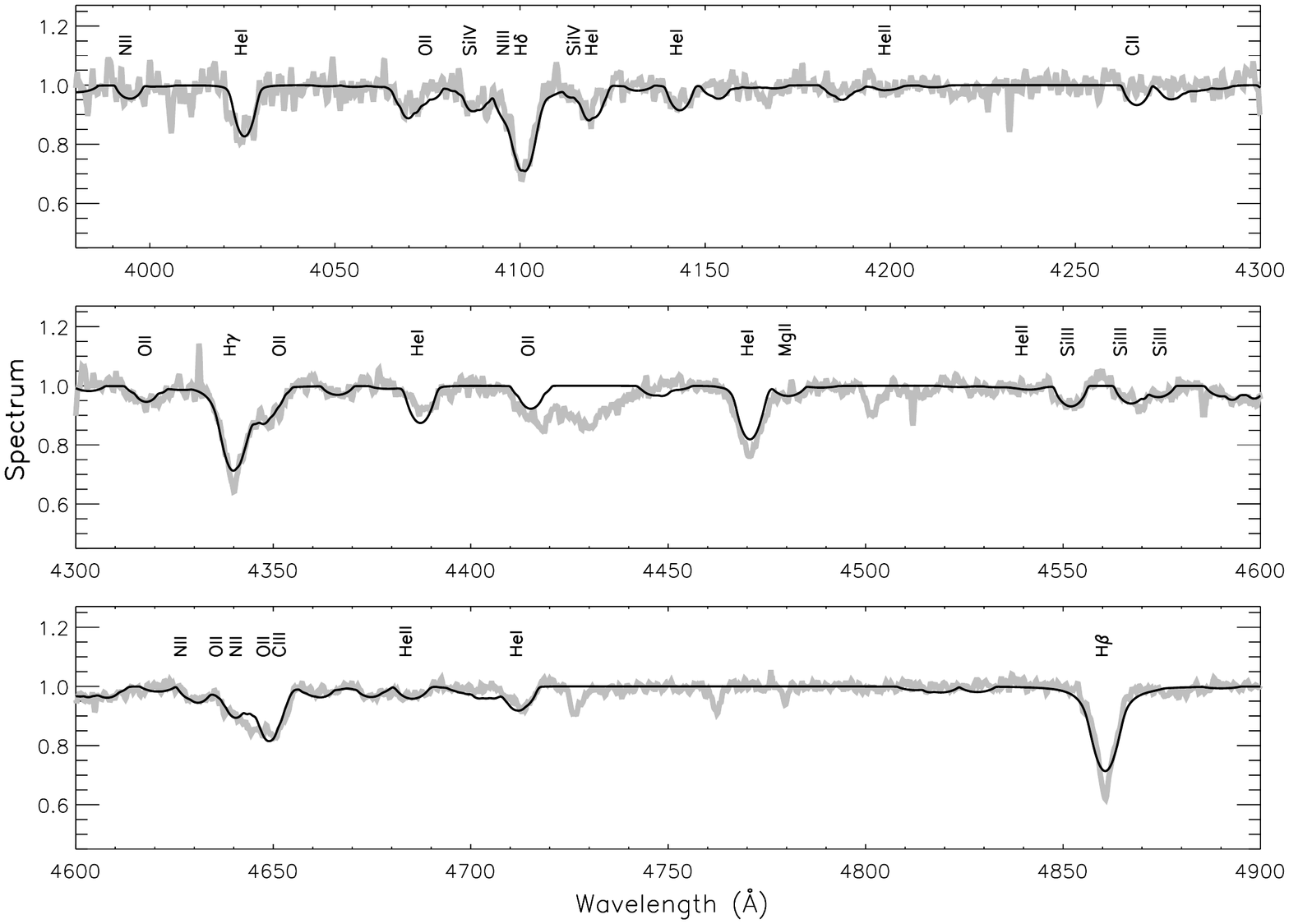}
\caption{Overplot of the X-shooter spectrum of $\igrjsqsc$ from 400 to 490\,nm with the stellar atmosphere model fit.}
\label{figure:modelisation}
\end{figure*}

A quantitative spectroscopic analysis was subsequently performed by means of \textsc{F}ast \textsc{A}nalysis of \textsc{ST}ellar atmospheres with \textsc{WIND}s \cite[\textsc{Fastwind};][]{santolaya:1997, puls:2005}, a spherical, non-LTE model atmosphere code with mass loss and line-blanketing that follows the strategy described in \cite{castro:2012}. It is based on an automatic $\chi^{2}$ fitting of synthetic \textsc{Fastwind} spectra including lines from \ion{H}, \ion{He}{i-ii}, \ion{Si}{ii-iv}, \ion{Mg}{ii}, \ion{C}{ii}, \ion{N}{ii-iii}, and \ion{O}{ii} to the global X-shooter spectrum between 390 and 510\,nm. The results of the fit are shown in Fig.~\ref{figure:modelisation}, and the stellar parameters and abundances measured from this fit are given in Tables~\ref{table:modelisation} and \ref{table:abundances}, respectively. 

While we find a velocity of $v = 320 \pm 8$\,$\mathrm{km}\,\mathrm{s}^{-1}$ from Fourier analysis, the line fitting favours a velocity of $v  = 280\:\mathrm{km}\,\mathrm{s}^{-1}$, obtained with errors of the order of 10\%, therefore likely consistent with $320 \pm 8\:\mathrm{km}\,\mathrm{s}^{-1}$. The stellar parameters -- effective temperature $T_{\mathrm{eff}} = 26000^{+1200}_{-1700}$~K and gravity $\log g = 3.10^{+0.15}_{-0.15}$-- are consistent with a star of spectral type B0.5\,Ib \citep{mcerlean:1999}, therefore consistent with the spectral classification described above. The helium abundance seems high, but the value obtained corresponds to the upper edge of the grid. However, we caution that the sensitivity of the He lines to abundance is not very high at this temperature. While Si, Mg, and O have solar abundances, the star seems mildly C- and N-enhanced within the large error bars\footnote{The large errors in the abundances are due to the signal-to-noise ratio, $v$, and the weakness of some key lines, which may blur the lines with the normalised continuum. Moreover, carbon and magnesium abundances are based only on one line each.}, but this is expected for a fast rotator.

\begin{table}
\caption{Stellar parameters measured from modelling X-shooter spectra with {\sc fastwind}. According to standard notation, $Y_{\rm He}$ is the abundance of He ratio ($\frac{N(He)}{N(H)}$). \label{table:modelisation}}
\scalebox{1.}{
\centering
 \begin{tabular}{lccc} 
 \hline
\noalign{\smallskip}
Parameter        & Value   & err-  & err+   \\
\noalign{\smallskip}
\hline
\noalign{\smallskip}
$T_{\mathrm{eff}}$ & 26000~K & -1700 & +1200  \\ 
$\log(g)$        & 3.10    & -0.15 & +0.15  \\
$\xi_{\mathrm{t}}$ & $13\:\mathrm{km}\,\mathrm{s}^{-1}$ & -2 & +2 \\
$Y_{\rm He}$      &  0.25   & $-0.1$ & $+0.1$ \\
\noalign{\smallskip}
\hline
\end{tabular}}
\end{table}

\begin{table}
\caption{Chemical abundances in $\log \left(\frac{N({\rm X})}{N({\rm H})}\right)+12$ resulting from the
{\sc fastwind} spectroscopic analysis. \label{table:abundances}}
\scalebox{1.}{
\centering
 \begin{tabular}{lccc} 
 \hline
\noalign{\smallskip}
Element & Abundance & err-  & err+ \\
\noalign{\smallskip}
 \hline
\noalign{\smallskip}
Si      & 7.6    & $-0.3$ & +0.3 \\ 
Mg      & 7.6    & $-0.7$ & +0.5 \\
C       & 8.8    & $-0.3$ & +0.2 \\ 
N       & 8.5    & $-0.2$ & +0.3 \\
O       & 8.7    & $-0.1$ & +0.1 \\
\noalign{\smallskip}
 \hline
\end{tabular}
} 
\end{table}

\subsection{Reddening correction}

To correct the magnitude from the reddening due to the absorption of the interstellar medium (ISM), we applied the same method described in \cite{chaty:2011a}. From the value of the column density $N(\mathrm{H}) = 8.7 \times 10^{21}\,\mathrm{cm}^{-2}$ \citep{nespoli:2008}, we calculated the corresponding colour excess $E(B-V)$ using \cite{bohlin:1978},

\begin{equation}
 N(H1+H2) = E(B-V) \times 5.8 \times 10^{21} \mbox{cm$^{-2}$ mag$^{-1}$}.
 \label{eq:EBV}
 \end{equation} 

We then determined the interstellar absorption in the $V$-band $\Av$,

 \begin{equation}
\Av = \Rv \times E(B-V) = 4.65,
\label{eq:Av}
\end{equation} 
where $R_{v} = 3.1$ represents the average extinction parameter in the Milky Way. 

Using the coefficients $a(x)$ and $b(x)$ given by \cite{cardelli:1989}, we derived A\textsubscript{$\lambda$}, the interstellar absorption in the other filters:
 
 \begin{equation}
 <A_{\lambda}/\Av > = a(x) + b(x)/ \Rv.
 \label{eq:cardcoeff}
  \end{equation} 

We finally subtracted these values from the apparent magnitudes for each wavelength:
 \begin{equation}
M_{\mathrm{dereddened}} = m_{\lambda} - A_{\lambda}.
\label{eq:fmder}
 \end{equation}
 
Following \cite{munari:2008}, another way to calculate $\Av$ is to determine $E(B-V)$ from the measurement of the equivalent width of the diffuse interstellar band (DIB) at 862.1\,nm (see X-shooter/VIS spectrum in Fig.~\ref{fig:xshvisspec}):
 
 \begin{equation}
E(B-V) = (27.2 \pm 0.3) \times EW(nm).
 \label{eq:Munform}
  \end{equation}
  
This DIB has been detected with $EW = 0.067$\,nm leading to a value of $E(B-V) =  1.82 \pm 0.02$. We thereafter apply the same process described above, and find $\Av = 5.65 \pm 0.06$.

\begin{figure}
\centering
\includegraphics[width=1.0\textwidth]{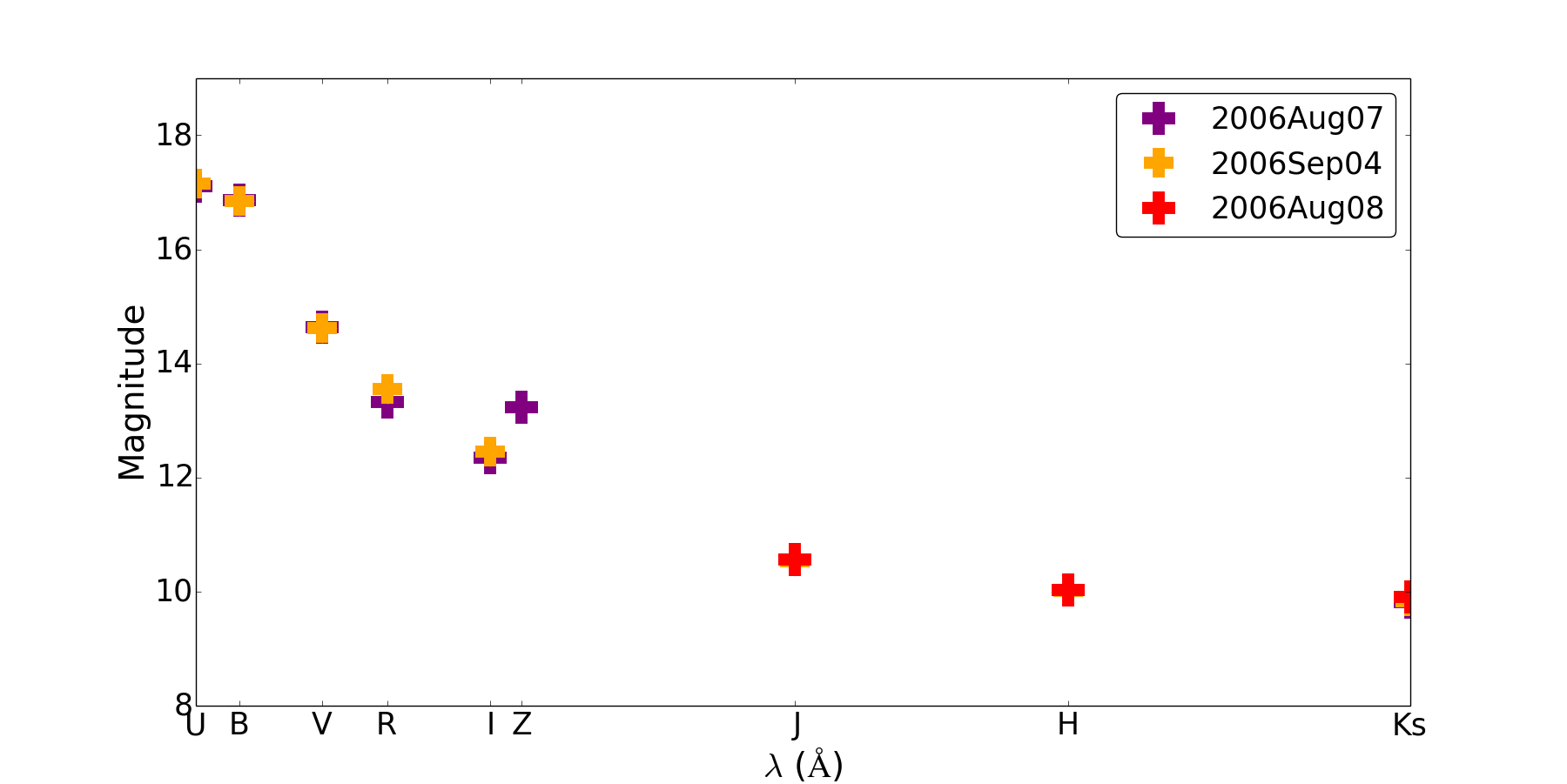}
\includegraphics[width=1.0\textwidth]{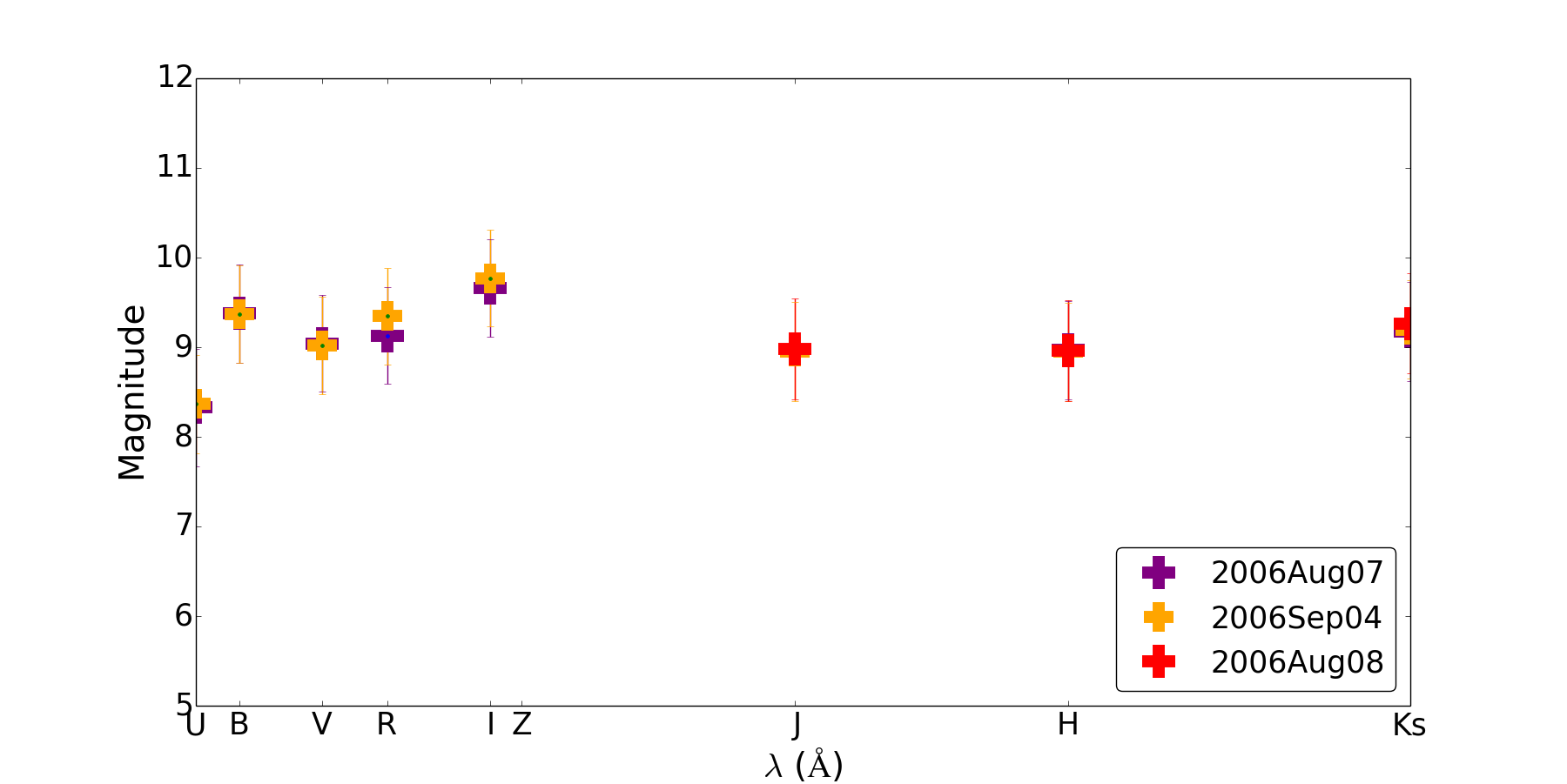}
\caption{Photometric magnitudes of $\igrjsqsc$, respectively before (upper panel) and after (lower panel) dereddenning.}
\label{fig:deredt}
\end{figure}

The value of $\Av = 4.65$ found with the photometric method is $\sim 1$\,mag smaller than the $\Av$ estimated by \cite{rahoui:2008a}, while the value of $\Av = 5.65 \pm 0.06$ estimated spectroscopically is well inside their confidence range. In the following, we use this latter value. As already noted by \cite{goldoni:2012}, this may suggest a non-standard extinction law in the direction of the source. We show in Fig.~\ref{fig:deredt} both reddened and dereddened magnitudes\footnote{As none of the coefficients was available for the $Z$ filter, the magnitude in this band is not dereddened.}.

\subsection{Spectral energy distribution}

\begin{figure}
\centering
\includegraphics[width=1.4\textwidth]{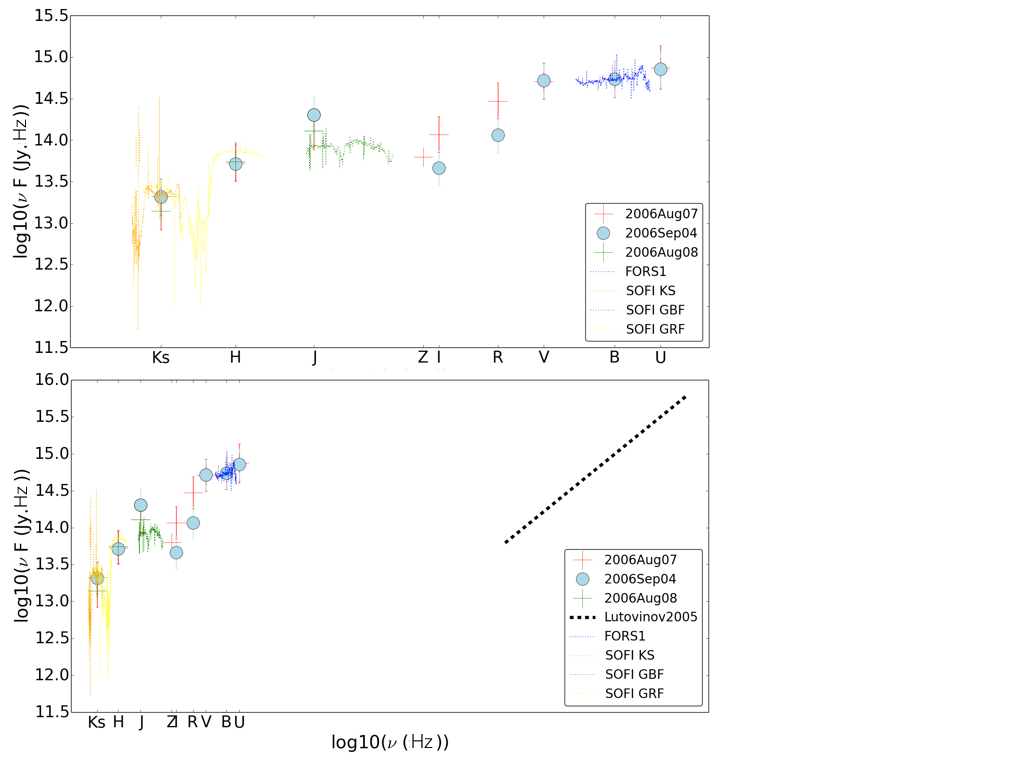}
\caption{Upper panel: OIR SED showing photometric and spectroscopic measurements. Lower panel: same SED including the 1-100 keV X-ray component.}
\label{fig:sed}
\end{figure}

We show in Fig.~\ref{fig:sed} (upper panel) the spectral energy distribution (SED) in the OIR domain. In addition to our photometric data, we included our SOFI and FORS1 spectra. The optical part represents a substantial contribution, consistent with the supergiant nature of the companion star. This SED presents a NIR excess which is characteristic of an envelope surrounding the supergiant star. 
 
In the lower panel of Fig.~\ref{fig:sed}, we also added the fitted X-ray component in the 1-100 keV energy band extracted from the work of \cite{lutovinov:2005a}. The X-ray part of the SED represents a contribution of roughly the same order as the optical part, favouring a scenario of stellar wind accretion, less luminous in X-rays than a direct accretion through an accretion disk \citep[see e.g.][]{tauris:2006}.
   
\subsection{Lightcurve}

\begin{figure}
\centering
\includegraphics[width=1.0\textwidth]{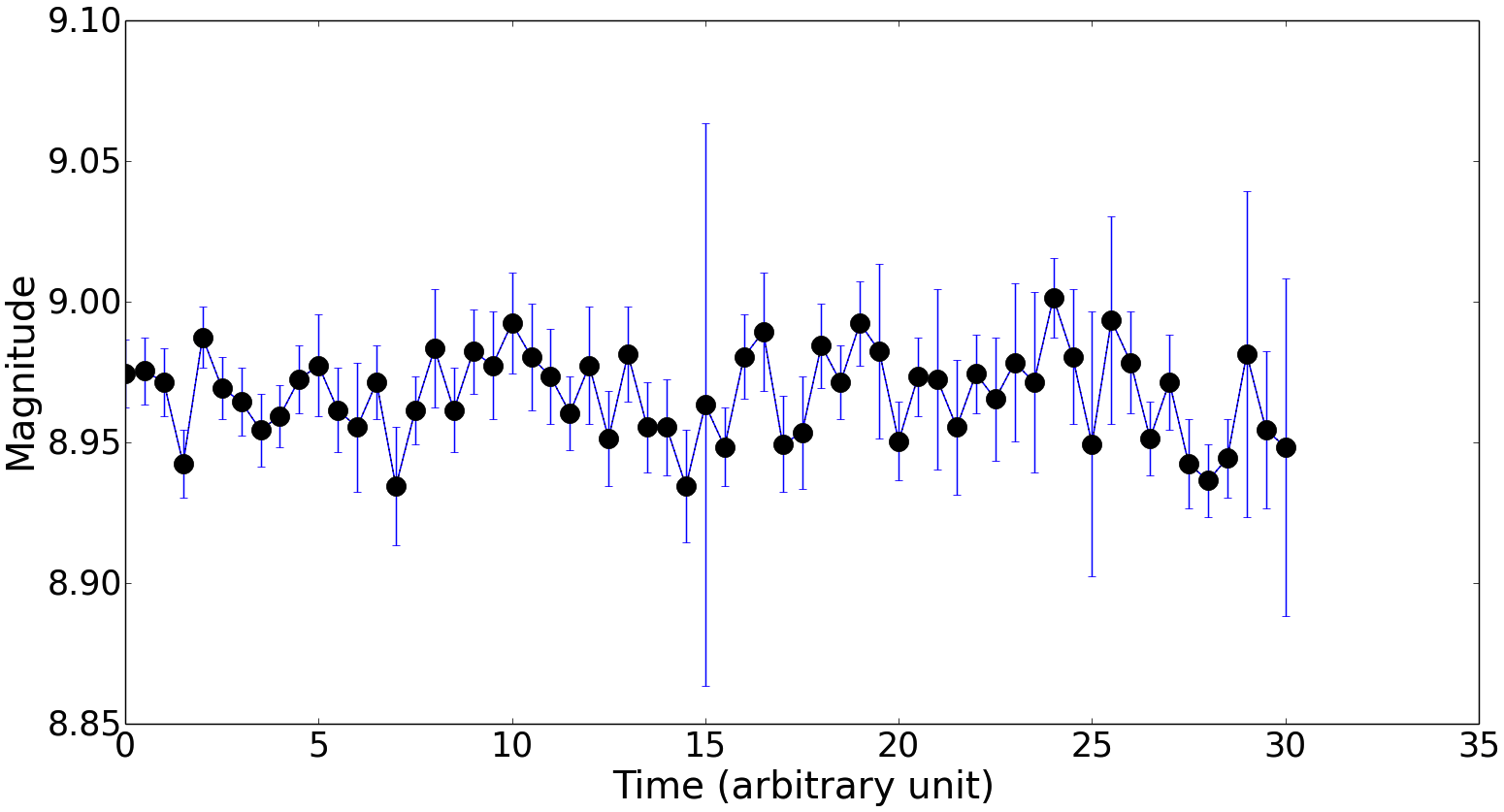}
\caption{Optical photometric lightcurve of $\igrjsqsc$ taken with SUSI2 in the V band.}
\label{fig:vlcurve}
\end{figure}

Short exposure images were acquired with SUSI2 in the $V$ band every 30\,s over 30\,min, on 2006 August 6 (Table~\ref{table:pgrm}). This rapid photometry (Fig.~\ref{fig:vlcurve}) in the optical allows us to study intrinsic luminosity variations on short timescales. While the variations we detect are of the same order as the error on the magnitudes, this result is consistent with micro-variability exhibited by supergiant stars of spectral types earlier than B \cite[see e.g. ][]{bresolin:2004,laur:2012}.

\subsection{Position in the Corbet diagram}

\begin{figure}
\centering
\includegraphics[width=1.1\textwidth]{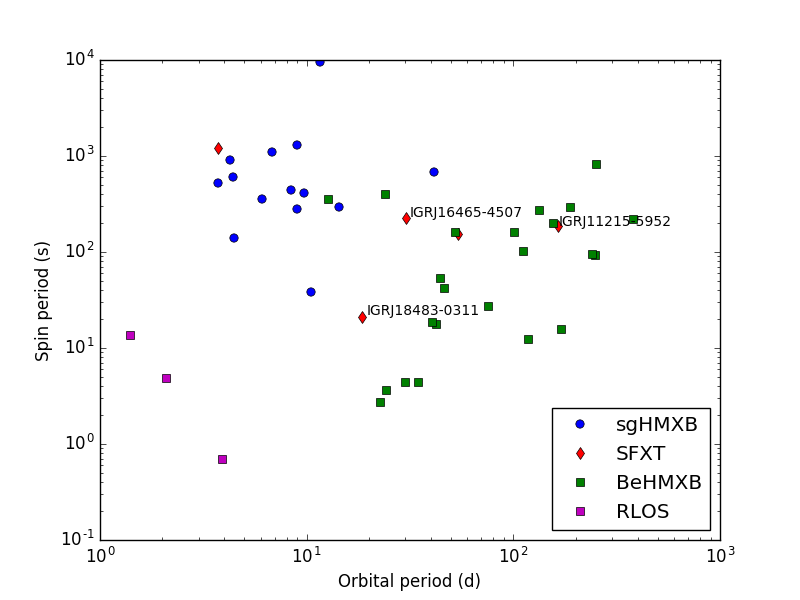}
\caption{Blue circles and green squares in this updated Corbet diagram represent wind-fed systems (sgHMXB) and BeHMXB, respectively. Red diamonds indicate SFXT; three are misplaced in this diagram: IGR\,J11215-5952, IGR\,J18483-0311, and $\igrjsqsc$. RLOS stands for Roche-lobe overflow systems.} 
\label{fig:corbetdiag}
\end{figure}

As described in Section \ref{section:introduction}, HMXB can be classified in three groups (sgHMXB or wind-fed systems, BeHMXB, and Roche-lobe filling systems) regarding their position in the orbital period versus spin period plot, the so-called Corbet diagram \citep{corbet:1986}. Contrary to the other two groups, BeHMXB exhibit a correlation between these two periods, which can be explained by transfer of angular momentum through accretion every time the compact object passes at periastron \cite[for more details see][]{chaty:2013}. From \cite{lutovinov:2005a} and \cite{clark:2010}, who respectively determined the spin ($228 \pm 6$~s) and the orbital period ($\sim 30$\,days), we present here an updated plot of this diagram (Fig.~\ref{fig:corbetdiag}) including HMXB from \cite{liu:2006} catalogue. 

This plot shows the peculiar position of $\igrjsqsc$, located right between the BeHMXB and sgHMXB area, even though our spectroscopic observations led us to conclude that its companion star was an early-B supergiant. Two other known SFXT (IGR\,J11215-5952 and IGR\,J18483-0311) have been found in the same area of the diagram and studied by \cite{liu:2011}, who suggested that some SFXT could be the descendants of BeHMXB. The high rotation velocity of the supergiant star might thus originate from the rapid rotator nature of BeHMXB.

While this rotation velocity is higher than the typical velocity of OB main sequence stars \cite[of the order of $\sim 150$\,km/s, see e.g.][]{murphy:2014}, it is not unlikely for a Be system. Taking an O7-8 main sequence star, its radius would increase by more than 50\% and its rotation velocity accelerate to $\sim 350$\,km/s, which approaches, but is still inferior to the break-up velocity \cite[close to $\sim 400$\,km/s, ][]{brott:2011}. Another possibility would be that the mass transfer from the primary has occurred in recent times, thus suggesting that we are facing a very young object.

\section{Conclusions} \label{section:conclusion}

We analysed all photometric and spectroscopic observations of $\igrjsqsc$ performed between 2006 and 2012, and taken from the ESO archive. From spectroscopic data, we constrain the companion star to be an early-B spectral type between B0.5 and B1\,Ib, in agreement with the preliminary determination by \cite{negueruela:2005a}.

The peculiar feature of this star, determined from spectroscopic observations and stellar fit modelling, is its high rotation velocity of $v = 320 \pm 8\:\mathrm{km}\,\mathrm{s}^{-1}$, suggesting a small radius of the supergiant star to avoid any disruption. We reported a SED in the OIR domain, showing a substantial contribution of the OIR emission from the companion star and an IR excess probably coming from a circumstellar envelope. We conclude that this system must be composed of a blue supergiant surrounded by a colder envelope, consistent with its SFXT classification as initially proposed by \cite{negueruela:2006a}. The group of sgHMXB of our Galaxy now includes 38 members, and $\igrjsqsc$ is the ninth member discovered by INTEGRAL with a confirmed spectral type \citep{coleiro:2013b}.

Furthermore, the position of this source on the Corbet diagram shows that it is probably a third member of SFXT located between Be and supergiant systems, and therefore possibly a descendant of BeHMXB. Finally, this system allows us to open new perspectives of research to study the evolution of HMXB in general, and the links between different populations of HMXB in particular.


\begin{acknowledgements}

We thank the anonymous referee for a constructive report. This work was supported by the Centre National d'Etudes Spatiales (CNES). It is based on observations obtained with MINE: the Multi-wavelength INTEGRAL NEtwork. This research is partially supported by the Spanish Mineco under grants AYA2012-39364-C02-01/02. The work of IND is also supported by the Spanish Ministerio de Educaci\'on y Ciencia under grant PRX14-00169. {\it IRAF} is distributed by the National Optical Astronomy Observatory, which is operated by the Association of Universities for Research in Astronomy (AURA) under a cooperative agreement with the National Science Foundation. This publication makes use of data products from the Two Micron All Sky Survey, which is a joint project of the University of Massachusetts and the Infrared Processing and Analysis Center/California Institute of Technology, funded by the National Aeronautics and Space Administration and the National Science Foundation. This research has made use of the USNO Image and Catalogue Archive operated by the United States Naval Observatory, Flagstaff Station (http://www.nofs.navy.mil/data/fchpix/).

\end{acknowledgements}


\input{chaty_igrj16465-final.bbl}

\clearpage
\begin{appendix}
\section{Identified lines}

\begin{table}
\caption{Line list in the FORS1 spectrum (Fig.~\ref{fig:FORS1optID}).\label{table:forsline}}
 \begin{center}
 \begin{tabular}{lr} 
 \hline
Element & $\lambda$ (nm) \\
 \hline
 H & 397.0\\
 \hline
  \ion{O}{ii} & 400.7\\ 
 \hline
 \ion{He}{i} & 402.6\\
\hline
 \ion{C}{iii} &  407.0 \\
 \hline
  H &  410.1 \\
 \hline
  \ion{He}{i} & 412.0 \\
 \hline
  H &  414.3 \\
 \hline
  H &  434.0 \\
 \hline
  \ion{He}{i} &  438.7 \\
 \hline
  \ion{He}{i} &  447.1 \\
 \hline
   \ion{O}{ii} &  455.2 \\
 \hline
   \ion{He}{i} & 465.0 \\
 \hline
   \ion{He}{i} &  471.3 \\
 \hline
\end{tabular}
\end{center}
\end{table}
\begin{table}
\caption{Line list in the EMMI spectrum (Fig.~\ref{fig:EMMIspec}).\label{table:emmiline}}
 \begin{center}
 \begin{tabular}{lr} 
 \hline
Element & $\lambda$ (nm) \\
 \hline
 \hline
  \ion{N}{ii} & 571.08 \\ 
 \hline
  \ion{C}{iv} & 581.20 \\ 
  \hline
  \ion{He}{i} & 587.56 \\ 
 \hline
  \ion{C}{ii} & 588.98 \\
 \hline
  \ion{N}{ii} & 594.16  \\
 \hline
  \ion{C}{i} & 601.32 \\ 
 \hline
  \ion{O}{i} & 604.64 \\
 \hline
  \ion{N}{ii} & 616.78 \\
 \hline
  \ion{O}{ii} & 619.79 \\
 \hline
  \ion{O}{i} & 637.43 \\
 \hline
  \ion{N}{i} & 644.00 \\ 
 \hline
  \ion{O}{ii} & 653.58 \\
 \hline
       H & 656.20 \\
 \hline
  \ion{C}{ii} & 657.80 \\
 \hline
  \ion{N}{ii} & 661.06 \\ 
 \hline
  \ion{He}{i} & 667.81 \\
 \hline
  \ion{O}{ii} & 688.49 \\
 \hline
  \ion{O}{ii} & 686.95 \\
 \hline
  \ion{He}{i} & 706.52 \\ 
 \hline
\end{tabular}
\end{center}
\end{table}
\begin{table}
\caption{Line list in the SOFI/GBF spectrum (Fig.~\ref{fig:SOFIgbfspec}).\label{table:gbfline}}
 \begin{center}
 \begin{tabular}{lr} 
 \hline
Element & $\lambda$ (nm) \\
 \hline
  \hline
  \ion{O}{i} &  950.56\\
 \hline
  \ion{He}{i} & 952.62 \\
  \hline
  H & 1004.98\\
  \hline
  \ion{O}{i} & 1042.12\\
  \hline
  H & 1093.82\\ 
 \hline
   \ion{O}{i} & 1128.63\\
 \hline
   \ion{O}{i} & 1135.87\\
 \hline
   \ion{He}{i} & 1508.37\\
 \hline
\end{tabular}
\end{center}
\end{table}  
\begin{table}
\caption{Line list in the SOFI/GRF (H part) spectrum (Fig.~\ref{fig:SOFIgrfspec_H}).\label{table:grfHline}}
 \begin{center}
 \begin{tabular}{lr} 
 \hline
Element &$\lambda$ (nm) \\
 \hline
 \hline
 \ion{H}{i} & 1641.0 \\
 \hline
  \ion{H}{i} & 1681.0 \\ 
 \hline
 \ion{He}{i} & 1700.0 \\
\hline
\ion{H}{i} &  1736.0 \\
 \hline
\end{tabular}
\end{center}
\end{table}  

\begin{table}
\caption{Line list in the SOFI/GRF (Ks part) spectrum (Fig.~\ref{fig:SOFIgrfspec_Ks}). \label{table:grfKsline}}
 \begin{center}
 \begin{tabular}{lr} 
 \hline
Element &$\lambda$ (nm) \\
 \hline
 \ion{He}{i} & 2058.69 \\ 
 \hline
  \ion{He}{i} & 2112.58 \\ 
 \hline
  H & 2166.12 \\ 
 \hline
\end{tabular}
\end{center}
\end{table}

\begin{table}
\caption{Line list in the X-shooter/UVB spectrum (Fig.~\ref{fig:xshuvbspec}).\label{table:xtabuvb}}
 \begin{center}
 \begin{tabular}{lr} 
 \hline
Element & $\lambda$ (nm) \\
\hline
  H\,(Ba\,SL) & 365.00 \\ 
\hline
 H\,(Ba\,9-2) & 383.54 \\ 
\hline
 H\,(Ba\,8-2) & 388.9 \\
\hline 
 H\,(Ba\,7-2) & 397.0 \\
\hline
 \ion{He}{i} & 402.62 \\ 
\hline
 \ion{O}{ii} & 407.00 \\ 
\hline
 H\,(Ba\,6-2) & 410.17 \\ 
\hline
 \ion{He}{i} & 412.08 \\ 
\hline
 \ion{He}{i} & 414.38 \\ 
\hline
 \ion{C}{ii} & 426.00 \\
\hline 
 \ion{O}{ii} & 431.71 \\
\hline
 H\,(Ba\,5-2) & 434.05 \\
\hline
 \ion{O}{ii} & 434.94 \\ 
\hline
 \ion{He}{i} & 438.79 \\ 
\hline
 \ion{O}{ii} & 441.50 \\ 
\hline
 \ion{He}{i} & 447.15 \\ 
\hline
 \ion{Mg}{ii} & 448.00 \\
\hline 
 \ion{Si}{iii} & 455.26 \\ 
\hline
 \ion{Si}{iii} & 456.78 \\ 
\hline
 \ion{O}{ii} & 464.91 \\
\hline 
 \ion{C}{iii} & 465.00 \\
\hline
 \ion{He}{i} & 471.31 \\ 
\hline
 H\,(Ba\,4-2) & 486.13 \\ 
\hline
 \ion{He}{i} & 492.19 \\ 
\hline
 \ion{N}{ii} & 500.51 \\ 
\hline
 \ion{He}{i} & 501.57 \\ 
\hline
 \ion{He}{i} & 504.77 \\ 
\hline
\end{tabular}
\end{center}
\end{table}

\begin{table}
\caption{Line list in the X-shooter/VIS spectrum (Fig.~\ref{fig:xshvisspec}). \label{table:xtabvis}}
 \begin{center}
 \begin{tabular}{lr}
\hline
Element     &$\lambda$ (nm) \\
\hline
\ion{N}{ii} & 566.66 \\ 
\hline
\ion{He}{i} & 587.56 \\ 
\hline
Na-D & 588.98 \\ 
\hline
Na-D & 589.52 \\ 
\hline
H\,(Ba\,3-2) & 656.30 \\ 
\hline
\ion{He}{i} & 667.82\\ 
\hline
\ion{He}{i} & 706.52 \\ 
\hline
\ion{He}{i} & 728.13 \\ 
\hline
\ion{O}{i} & 777.19 \\ 
\hline
 H\,(Pa\,SL) & 820.00 \\
\hline 
H\,(Pa\,20-3) & 839.24\\
\hline
H\,(Pa\,19-3) & 841.33 \\
\hline
H\,(Pa\,18-3) & 843.80 \\
\hline
H\,(Pa\,17-3) & 846.73 \\
\hline
H\,(Pa\,16-3) & 850.25 \\
\hline
H\,(Pa\,15-3) & 854.54\\
\hline
H\,(Pa\,14-3) & 859.84 \\
\hline
H\,(Pa\,13-3) & 866.50 \\
\hline
H\,(Pa\,12-3) & 875.05 \\
\hline
H\,(Pa\,11-3) & 886.28 \\
\hline
H\,(Pa\,10-3) & 901.53 \\
\hline
H\,(Pa\,9-3) & 922.97 \\
\hline
H\,(Pa\,8-3) & 954.62 \\
\hline
H\,(Pa\,7-3) & 1004.98 \\
\hline
\end{tabular}
\end{center}
\end{table}

\begin{table}
\caption{Line list in the X-shooter/NIR spectrum (Fig.~\ref{fig:xshnirspec}). \label{table:xtabnir}}
 \begin{center}
 \begin{tabular}{lr} 
 \hline

 Element       & $\lambda$ (nm) \\
 \hline
 H\,(Pa\,7-3)  & 1004.98 \\ 
 \hline
 \ion{He}{ii}  & 1012.36 \\
 \hline
 \ion{He}{i}   & 1031.12 \\
 \hline
 \ion{He}{i}   & 1083.03 \\
 \hline
 \ion{He}{i}   & 1091.30 \\
 \hline 
 H\,(Pa\,6-3)  & 1093.82 \\ 
 \hline
 \ion{He}{i}   & 1196.91 \\ 
 \hline
 \ion{He}{i}   & 1252.75 \\
 \hline
 \ion{He}{i}-D & 1278.48 \\ 
 \hline
 H\,(Pa\,5-3)  & 1281.81 \\ 
 \hline 
 \ion{He}{i}   & 1296.84 \\ 
 \hline
 \ion{He}{i}   & 1508.36 \\ 
 \hline
 H\,(Br\,19-4) & 1526.50 \\
 \hline
 H\,(Br\,18-4) & 1534.60 \\
 \hline
 H\,(Br\,17-4) & 1544.30 \\
 \hline
 H\,(Br\,16-4) & 1556.05 \\ 
 \hline 
 H\,(Br\,15-4) & 1570.50 \\
 \hline
 H\,(Br\,14-4) & 1588.50 \\
 \hline
 H\,(Br\,13-4) & 1611.40 \\
 \hline
 H\,(Br\,12-4) & 1641.14 \\ 
 \hline
 H\,(Br\,11-4) & 1681.11 \\ 
 \hline
 \ion{He}{i}   & 1700.25 \\ 
 \hline
 H\,(Br\,10-4) & 1736.69 \\ 
 \hline
 H\,(Br\,9-4)  & 1818.10 \\
 \hline
 H\,(Pa\,4-3)  & 1875.6 \\ 
\hline 
 H\,(Br\,8-4)  & 1945.10 \\
 \hline
 \ion{He}{i}   & 2058.69 \\ 
 \hline
 \ion{He}{i}   & 2112.58 \\ 
 \hline
 H\,(Br\,7-4)  & 2166.12 \\ 
 \hline
 H\,(Pf\,SL)  &  2279.0 \\
\hline
\end{tabular}
\end{center}
\end{table}

\end{appendix}


 \end{document}

%% file: symbols.tex
\def\igrjsqsc{\mbox{IGR~J16465$-$4507}}

\def\microns{\mbox{ } \mu \mbox{m}}

\def\adeg{^{\circ}}

\def\asec{^{\prime \prime}}

\def\Av{A_{\rm V}}

\def\Rv{R_{\rm v}}
\def\ltsima{\; \buildrel < \over \sim \;}
\def\simlt{\lower.5ex\hbox{\ltsima}}            
\def\simgt{\lower.5ex\hbox{\gtsima}}            
\def\gtsima{\; \buildrel > \over \sim \;}